\documentclass[fleqn,10pt]{wlscirep}
\usepackage[utf8]{inputenc}
\usepackage{float}
\usepackage{diagbox}
\usepackage[T1]{fontenc}
\usepackage{subcaption}

\title{Decoding Selective Auditory Attention to Musical Elements in Ecologically Valid Music Listening}

\usepackage{parskip}
\author[1,+,*]{Taketo Akama}
\author[1,+]{Zhuohao Zhang}
\author[1]{Tsukasa Nagashima}
\author[1]{Takagi Yutaka}
\author[1]{Shun Minamikawa}
\author[1]{Natalia Polouliakh}

\affil[1]{Sony Computer Science Laboratories, Inc, Tokyo, Japan}
\affil[*]{Corresponding author: Taketo Akama, taketo.akama@gmail.com}
\affil[+]{these authors contributed equally to this work}

\begin{abstract}
Art has long played a profound role in shaping human emotion, cognition, and behavior. While visual arts such as painting and architecture have been studied through eye-tracking, revealing distinct gaze patterns between experts and novices, analogous methods for auditory art forms remain underdeveloped. Music, despite being a pervasive component of modern life and culture, still lacks objective tools to quantify listeners' attention and perceptual focus during natural listening experiences.  
To our knowledge, this is the first attempt to decode selective attention to musical elements using naturalistic, studio-produced songs and a lightweight consumer-grade EEG device with only four electrodes. By analyzing neural responses during real-world–like music listening, we test whether decoding is feasible under conditions that minimize participant burden and preserve the authenticity of the musical experience. 
Our contributions are fourfold:
(i) decoding music attention in real studio-produced songs,
(ii) demonstrating feasibility with a four-channel consumer EEG,
(iii) providing insights for music attention decoding, and
(iv) demonstrating improved model ability over prior work.
Our findings suggest that musical attention can be decoded not only for novel songs but also across new subjects, showing performance improvements compared to existing approaches under our tested conditions. 
These findings show that consumer-grade devices can reliably capture signals, and that neural decoding in music could be feasible in real-world settings. This paves the way for applications in education, personalized music technologies, and therapeutic interventions.

\end{abstract}
\begin{document}

\flushbottom
\maketitle
\thispagestyle{empty}

\section*{Introduction}
Music is one of the most universally experienced and emotionally powerful forms of human expression. Across cultures and contexts, it engages perceptual, affective, and cognitive systems in ways that are both deeply personal and socially shared. Despite its ubiquity, our scientific understanding of how listeners attend to specific elements within complex musical scenes remains limited. In other sensations such as visual domains, perceptual focus can be tracked directly through techniques like eye-tracking, enabling detailed comparisons of attentional behavior between novices and experts. In contrast, auditory domains—particularly music—lack analogous tools for objectively measuring attentional allocation. Unlike visual attention, which can be directly inferred from observable eye movements, auditory attention involves internal neural processes with no overt behavioral correlates. All auditory information is encoded and processed within the brain, primarily in the auditory cortex, making it inherently more challenging to detect and interpret from external measurements. Moreover, the neural representation of competing auditory streams remains only partially understood, further complicating efforts to track how listeners allocate attention in complex acoustic environments. This gap has hindered progress in understanding how the brain dynamically selects, filters, and engages with musical components during natural listening. Addressing this challenge is essential not only for advancing the neuroscience of music cognition, but also for enabling new technologies that adapt to the listener’s perceptual state in real time.

Understanding how the human brain selectively attends to different auditory elements is a fundamental question in neuroscience. Although significant progress has been made in decoding auditory attention in speech-based paradigms—particularly in the classic “cocktail party” scenario\cite{19,20,21,27,28,29,30,34,35,36}, these studies primarily focus on linguistic content and directional cues, which are not necessarily central to musical listening. Some studies have investigated music-related attention using simple tones or isolated instrument streams\cite{18,24,26,32,33,39}, or pure tones\cite{25}, rather than naturalistic music. Furthermore, many of these studies rely on spatial listening paradigms, where different audio streams are presented separately to each ear or from distinct spatial locations\cite{17,18,24,26}, thereby confounding attentional focus with spatial cueing. In addition to music, there are BCI studies that introduce tactile stimuli as an additional sensory modality, where attention is distinguished between the left and right ears for music and between the left and right hands for tactile stimuli; however, this is also related to spatial attention\cite{23}. 

Most studies using real music focus on analyzing the brain’s functional responses to auditory stimuli under natural listening conditions. These works typically examine neural tracking of musical features such as onset and tempo\cite{6,8,9,11}, individual differences in EEG responses\cite{4}, or comparisons between musicians and non-musicians \cite{3}. Others focus more broadly on how music affects brain states from a neuroscientific perspective\cite{5,7,10,12}. Additional research explores the oddball paradigm in music\cite{42}, musical imagery\cite{9,44} or contrasts between speech and music\cite{22}. However, to our knowledge, no work has attempted to decode auditory attention in the context of studio-produced music characterized by real instrumentation, complex production, and diverse genres.

Among existing works, the study most relevant to ours is the study\cite{39}, which investigated attention to one instrument within duos or trios from the MAD-EEG dataset. In their work, each instrument track was recorded in mono, and the stereo condition was created by assigning these mono tracks to different spatial positions through conventional panning. By contrast, our study used studio-produced songs in which each instrument naturally contributes to a two-channel stereo mix, thereby preserving the authentic spatial characteristics of real music. In addition, the audio stimuli in their work were synthesized compositions limited to only two or three elements, whereas real-world pop music typically contains many more instrumental layers and complex production. Taken together, these differences highlight a substantial gap between prior experimental stimuli and the ecological richness of real studio-produced music, which our study aims to bridge.

Taken together, while auditory attention decoding has advanced considerably in controlled speech or tone-based contexts, research on attention in real-world music remains limited. Existing studies often rely on spatial or acoustic simplifications, and few have addressed the challenge of decoding attention in complex, polyphonic, and genre-diverse music. Moreover, prior work has largely focused on characterizing neural responses rather than developing systems or applications that leverage decoded attentional states. The incorporation of real-world recordings is increasingly recognized as essential for advancing auditory attention research\cite{realworld}. Yet, progress in this direction has been slow, as capturing neural responses to music under ecologically valid conditions is particularly challenging due to the complex social and emotional factors involved\cite{realworlddifficulty}. This gap strongly motivates the present study. Therefore, we propose to decode selective attention to musical components (Vocal, Drum, Bass, Others) in real-world, studio-produced songs, using only four channels of consumer-grade EEG for practical scalability.

To our knowledge, this is the first study to investigate auditory attention decoding using actual studio-produced songs—not just simplified instrument tracks or synthetic sound mixtures. The recordings feature real instrumentation, professional production, and genre diversity, and the data were collected in a listening context closely resembling real-world music consumption. Rather than relying on spatial cues or deviant event detection, we decode listeners’ selective attention to specific musical components, such as vocal, drum, bass and others. This paradigm more closely resembles everyday music listening and represents a meaningful shift from artificial laboratory conditions to ecologically valid, application-relevant settings. Additionally, we are the first to explicitly examine the role of musical genre (e.g., pop, rock, jazz, electronic) as a variable in attention decoding. By analyzing EEG responses to genre-diverse songs, our study investigates whether decoding performance generalizes across musical styles—bridging the gap between controlled paradigms and the richness of real-world music perception. 

Furthermore, we employ the Muse2\cite{muse2}, a lightweight consumer-grade EEG device with dry electrodes, which substantially reduces setup time and imposes minimal user burden. 
Unlike laboratory-grade systems, consumer-grade EEG devices—such as Muse2 and other simplified headsets\cite{simpleEEG} enable neural recordings under conditions that more closely approximate real-world music listening. The use of this device offers several advantages: (i) it increases the ecological validity of our findings, facilitating generalization to everyday listening contexts; (ii) it lowers the technical and logistical barriers for conducting naturalistic studies, thereby accelerating scientific progress; and (iii) it simplifies data acquisition and experimental setup, making large-scale, real-world investigations of music cognition more feasible.

This research is driven not only by theoretical interest in music cognition and neural decoding, but also by its potential for broad real-world applications. By visualizing which musical components a listener attends to using EEG, we can reconstruct attended elements and generate personalized music. This can be applied in personalized music recommendation, music education (e.g., training selective listening skills), or wellness interventions that promote mindful and attentive music listening for mental health support and personalized therapeutic use. By bridging the gap between neural decoding and ecologically valid music listening, this study contributes to laying a foundation for neuroadaptive auditory technologies—advancing both fundamental understanding of music cognition and the development of brain-driven, listener-centric musical interfaces.

In summary, our work makes four key contributions:

(i) Ecological paradigm: We directly decode selective attention to specific musical components (Vocal/Drum/Bass/Others) in studio-produced, polyphonic songs without artificial spatial separation, enhancing ecological validity.

(ii) Practical Feasibility: We demonstrate stable performance with a four-channel consumer-grade EEG headset (Muse2), highlighting its suitability for large-scale and real-world use, while also lowering the barrier to data collection in related research.

(iii) Electrode configuration insight: Our findings suggest that a frontal–temporal four-electrode montage can effectively support decoding of selective musical attention, and may provide a useful basis for exploring how such configurations relate to established neuroscientific views on attention and auditory processing.

(iv) Generalization and comparison: Our results demonstrate robust decoding accuracy across both unseen-song/within-subject and unseen-subject/within-song settings, consistently outperforming the strongest available baseline identified in recent survey work, which performs classification without relying on artificial spatial information and is therefore representative of real-world music listening scenarios.

\section*{Results}
\subsection*{New songs evaluation within subject}
\subsubsection*{Global evaluation}
\begin{figure}[ht]
\centering
\includegraphics[width=0.6\linewidth]{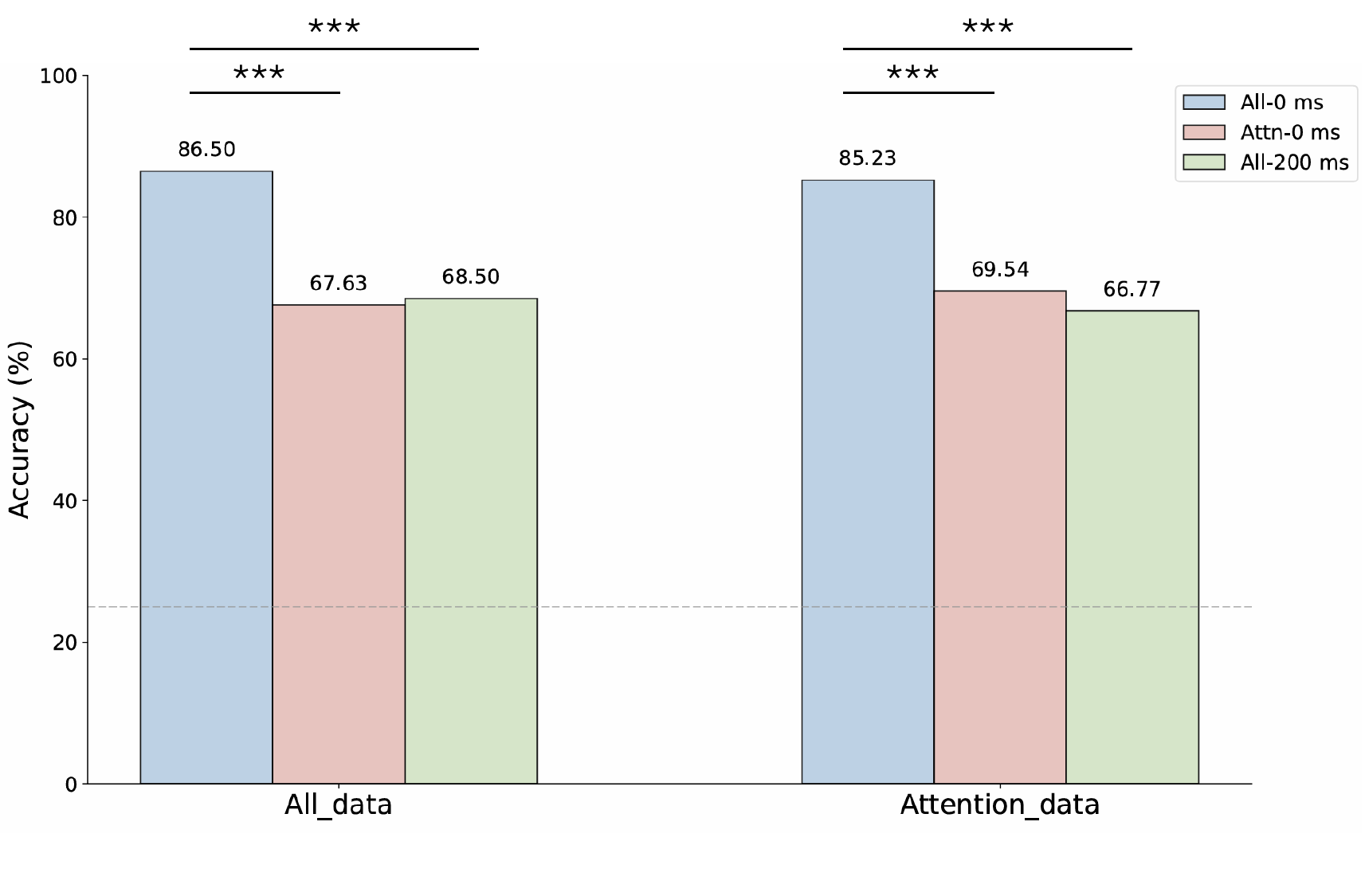}
\caption{Global accuracy.}
\label{fig:accuracy}
\begin{minipage}{0.8\textwidth}
\normalsize
Overall accuracies of the three models were calculated for both all-data and high-attention data evaluations. \textbf{Model: all-0 ms} showed highest accuracy than other two models. $^{*} p < 0.05$, $^{**} p < 0.01$, $^{***} p < 0.001$ (McNemar test, compared between different baseline models).
\end{minipage}
\end{figure}

We first evaluated the global accuracy of the three baseline models to summarize their overall accuracy.
Two baseline models were evaluated first: \textbf{Model: all-0 ms}, trained on the full trials without delay between EEG and audio signals; \textbf{Model: attn-0 ms}, trained only on high-attention trials (self-reported scores 4–5). 
The better-performing model was subsequently extended with a fixed 200-ms EEG–audio delay, motivated by prior work \cite{predann}, which we refer to as \textbf{Model: all-200 ms}.
Figure \ref{fig:accuracy} illustrates the results, with the x-axis divided into two groups: evaluation on all data and evaluation on high-attention data only. The three models are shown in different colors: \textbf{Model: all-0 ms} in blue, \textbf{Model: attn-0 ms} in red, and \textbf{Model: all-200 ms} in green.

The global accuracy is computed as the proportion of cases in which the similarity of a positive pair exceeds those of all three negative pairs simultaneously (e.g., 1-vs-3 comparison). Moreover, because the number of trials is unbalanced across tasks---with \textit{Vocal} accounting for 31 out of 63 trials---the global accuracy is more strongly weighted by the performance on the \textit{Vocal} task.

In terms of global accuracy, \textbf{Model: all-0 ms} achieved the highest performance, surpassing \textbf{Model: attn-0 ms} and \textbf{Model: all-200 ms} by 18.87\% and 18.00\% in the all-data evaluation. In the high-attention evaluation, \textbf{Model: all-0 ms} also maintained its superiority, outperforming the two models by 15.69\% and 18.46\%, respectively. To statistically evaluate differences in classification performance between models, we conducted pairwise McNemar tests on trial-wise correctness (correct vs.\ incorrect). \textbf{Model: all-0 ms} significantly outperformed both \textbf{Model: attn-0 ms} and \textbf{Model: all-200 ms} in the all-data evaluation (McNemar test, $p = 1.89 \times 10^{-25}$ and $p = 2.31 \times 10^{-19}$, respectively). A similar pattern was observed in the high-attention evaluation ($p = 2.57 \times 10^{-16}$ and $p = 9.76 \times 10^{-17}$, respectively).

\begin{figure}[H]
\centering
\includegraphics[width=0.6\linewidth]{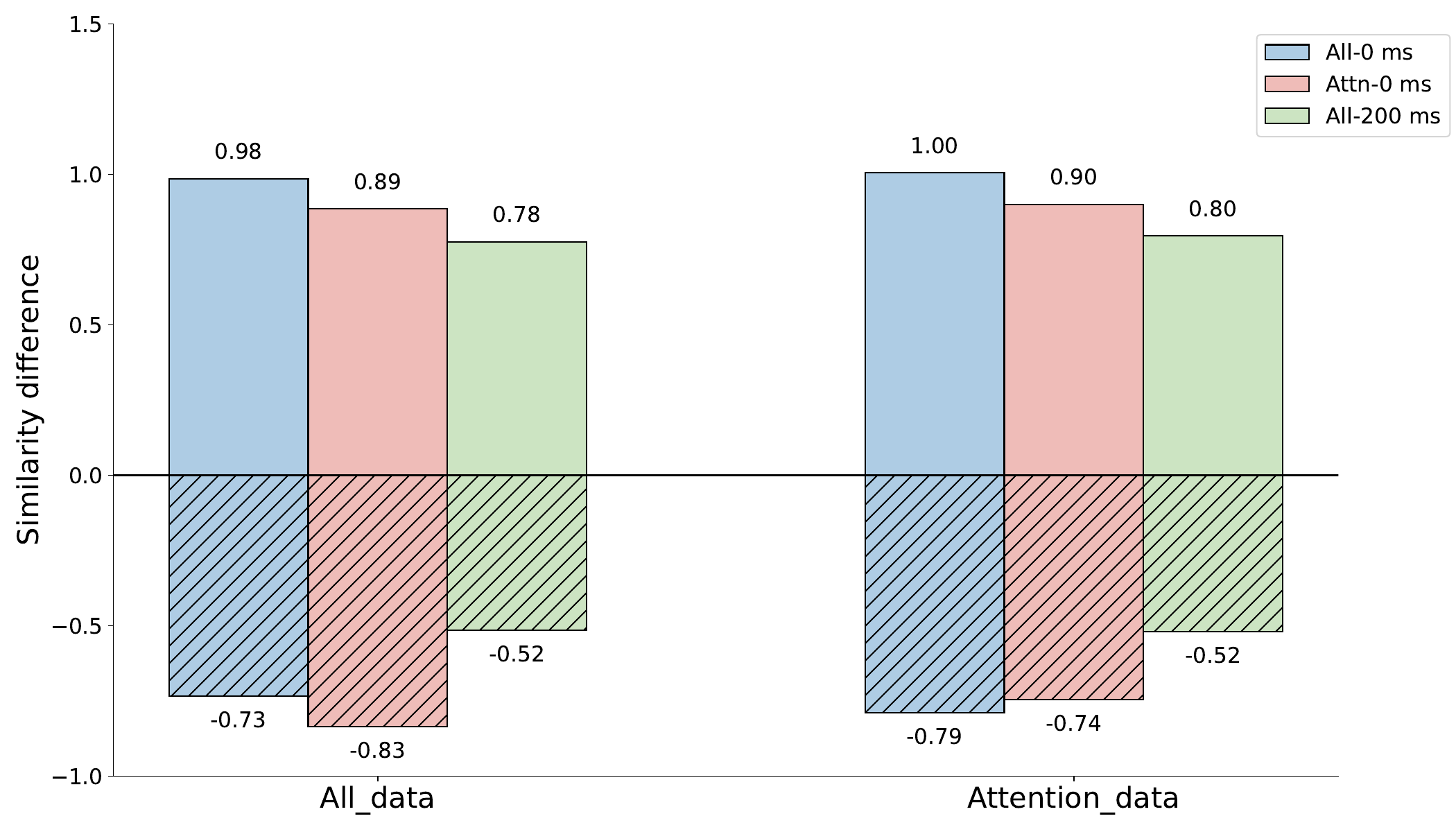}
\caption{Global similarity differences.}
\label{fig:difference}
\begin{minipage}{0.8\textwidth}
\normalsize
The graph is divided into two parts: positive bars above zero represent the \textbf{Correctly Matched} group, while shaded bars below zero correspond to the \textbf{Incorrectly Matched} group. Different colors indicate different models. For the \textbf{Correctly Matched} group, \textbf{Model: all-0 ms} yielded the largest similarity differences compared to the other two models.
\end{minipage}
\end{figure}

We further analyzed absolute global similarity values difference. The full test dataset was divided into two groups:
\textbf{Correctly Matched} group (positive pair similarity > all negatives) and an  \textbf{Incorrectly Matched} group (positive pair $\leq$ at least one negative), consistent with the criterion used for the global accuracy. This approach allows finer-grained inspection of model behavior.
Global difference results are shown in Figure \ref{fig:difference}, where the x-axis is divided into two groups: evaluation on all test data and evaluation on high-attention data. In the figure, positive bars above zero on the y-axis represent the \textbf{Correctly Matched} group, while shaded bars below zero correspond to the \textbf{Incorrectly Matched} group. The three models are indicated by different colors: blue for \textbf{Model: all-0 ms}, red for \textbf{Model: attn-0 ms}, and green for \textbf{Model: all-200 ms}.

For the \textbf{Correctly Matched} group, \textbf{Model: all-0 ms} exhibited the largest difference, followed by \textbf{Model: attn-0 ms}, while \textbf{Model: all-200 ms} showed the smallest. This pattern indicates that \textbf{Model: all-0 ms} provides the strongest discrimination between tasks when the model’s predictions are correct. By contrast, within the \textbf{Incorrectly Matched} group, smaller similarity differences would typically indicate better separation. Although \textbf{Model: all-0 ms} displayed larger similarity differences than \textbf{Model: all-200 ms} in the all-data evaluation and the largest differences overall in the high-attention evaluation, this likely reflects the reduced number due to its high accuracy, that only the most challenging cases remained in \textbf{Incorrectly matched} group. 
Therefore, the results may reflect this selection bias rather than a genuine deficit in the model’s ability to discriminate between tasks.

Overall, in the global evaluation, \textbf{Model: all-0 ms} outperformed the other two models.

\subsubsection*{Task-level evaluation}
\begin{figure}[htbp]
\centering

\begin{subfigure}{0.3\textwidth}
    \centering
    \includegraphics[width=\linewidth]{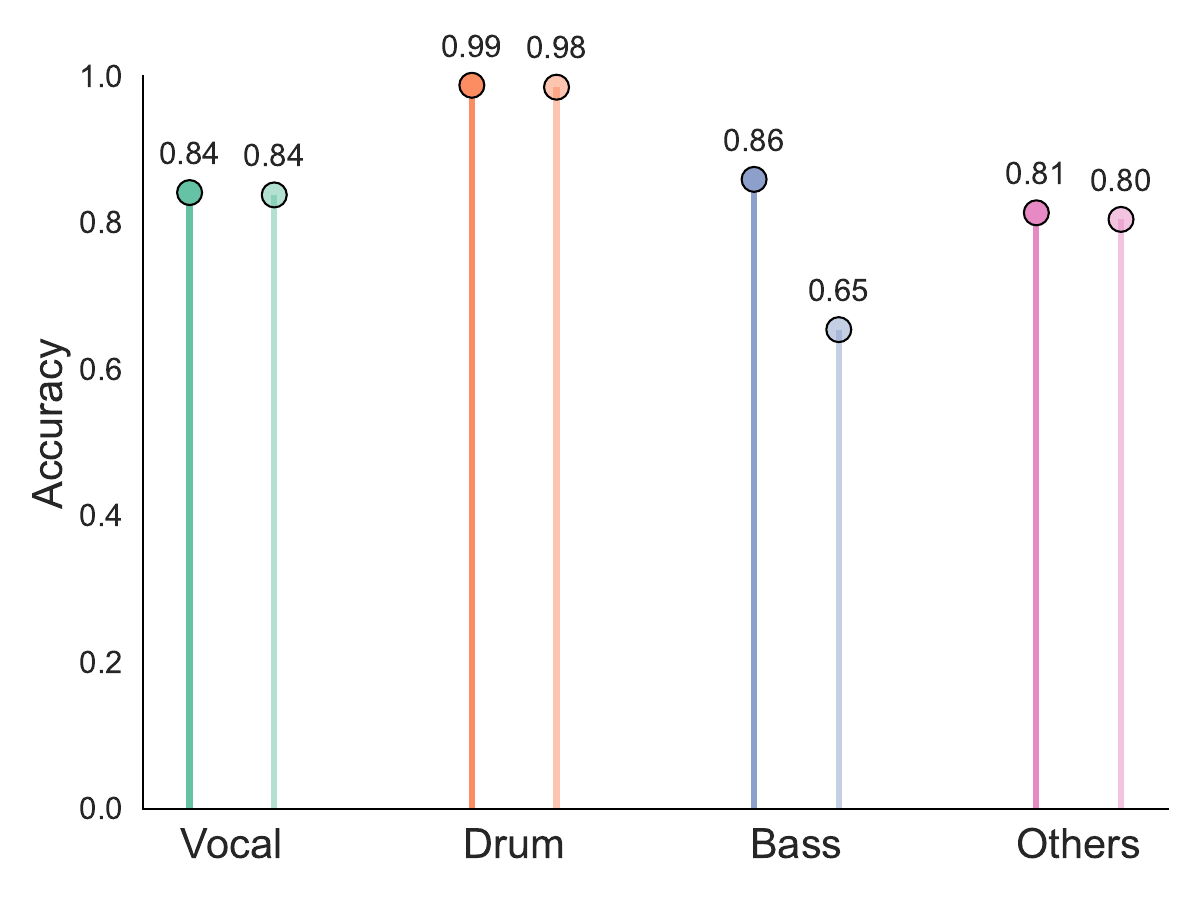}
    \caption{All-0 ms}
    \label{fig:subfig1}
\end{subfigure}
\begin{subfigure}{0.3\textwidth}
    \centering
    \includegraphics[width=\linewidth]{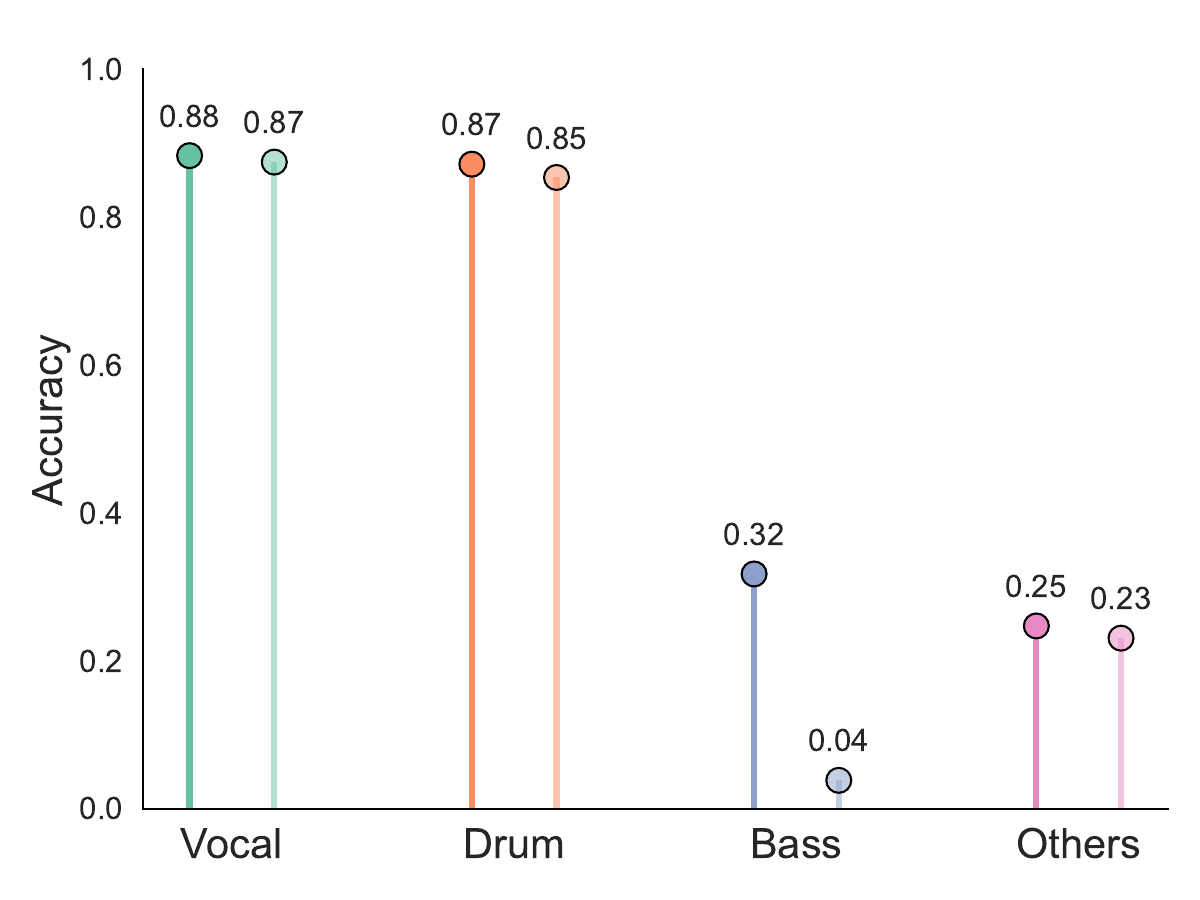}
    \caption{Attn-0 ms}
    \label{fig:subfig2}
\end{subfigure}
\begin{subfigure}{0.3\textwidth}
    \centering
    \includegraphics[width=\linewidth]{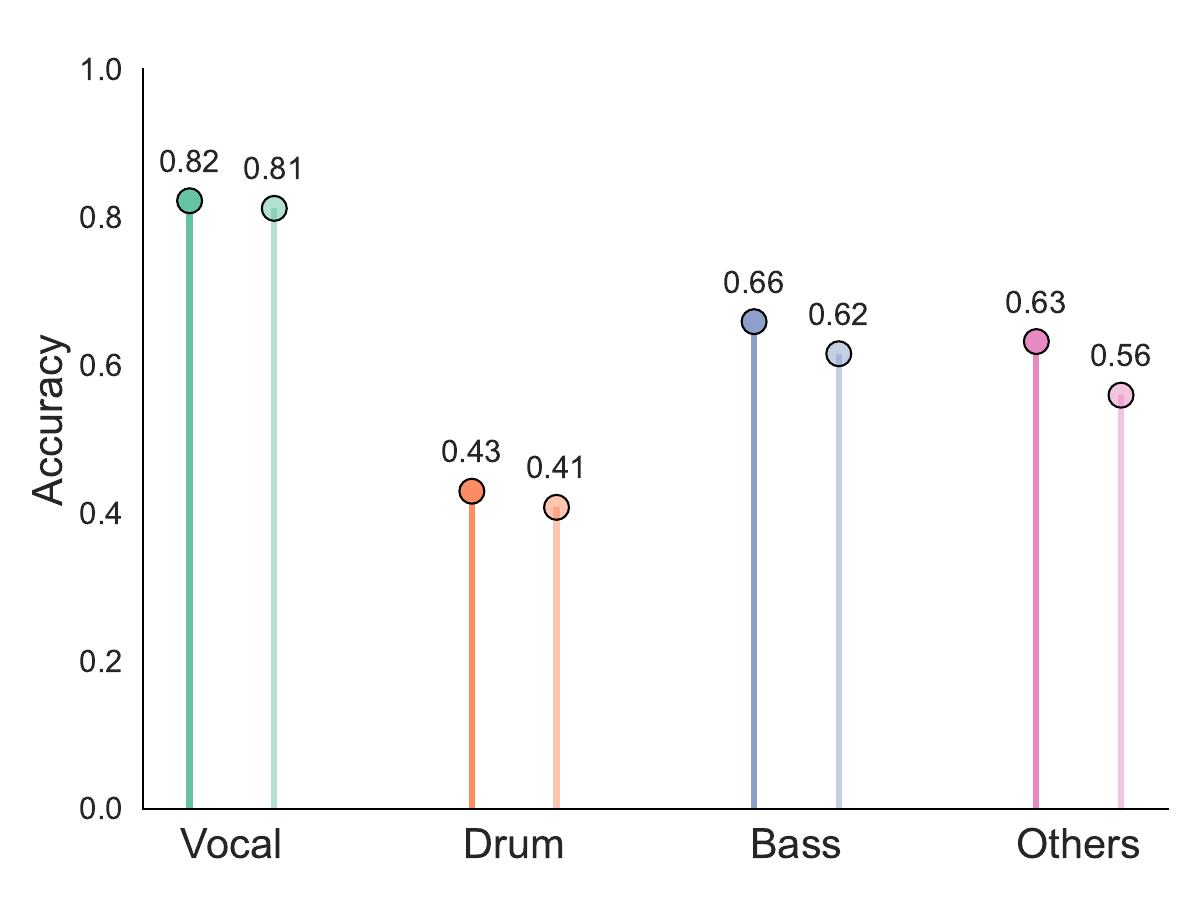}
    \caption{All-200 ms}
    \label{fig:subfig3}
\end{subfigure}

\caption{Task-level accuracy vectors.}
\label{fig:4-vector}
\begin{minipage}{0.8\textwidth}
\normalsize
Evaluation on all data is indicated by dark colors, whereas evaluation restricted to high-attention data is indicated by lighter colors. \textbf{Model: all-0 ms} showed relatively high accuracy across all tasks, while \textbf{Model: attn-0 ms} exhibited significant drops in \textit{Bass} and \textit{Others}, especially when focusing on high-attention data.
\end{minipage}
\end{figure}

We then computed a 4$\times$1 vector of task-level accuracy to represent performance for each task. Figure \ref{fig:4-vector} shows the results with solid lines represent evaluation on all test data, whereas lighter-colored lines indicate evaluation restricted to high-attention data. Colors denote different tasks: green for \textit{Vocal}, orange for \textit{Drum}, blue for \textit{Bass}, and pink for \textit{Others}.

\textbf{Model: all-0 ms} achieved over 80\% accuracy across all tasks in the all-data evaluation. Except for the \textit{Bass} task, which reached 65\%, all other tasks exceeded 80\% accuracy in high-attention data evaluation. In contrast, \textbf{Model: attn-0 ms} demonstrated comparable accuracy in the \textit{Vocal} and \textit{Drum} tasks but showed substantial performance drops in the \textit{Bass} and \textit{Others} tasks. This decline was particularly pronounced in the high-attention data evaluation, where nearly all \textit{Bass} trials were incorrectly matched (0.04\%). This may be due to the limited number of trials entering the high-attention dataset for the \textit{Bass} and \textit{Others} tasks, as participants self-reported greater difficulty maintaining attention during these tasks. \textbf{Model: all-200 ms} showed similar in \textit{Vocal}, but declines in other three tasks compared to \textbf{Model: all-0 ms} across both all-data and high-attention data evaluations. 
These 4$\times$1 vector of task-level accuracy indicate that \textbf{Model: all-0 ms} appears to be the most robust among the tested models.

\begin{figure}[htbp]
\centering

\begin{subfigure}{0.3\textwidth}
    \centering
    \includegraphics[width=\linewidth]{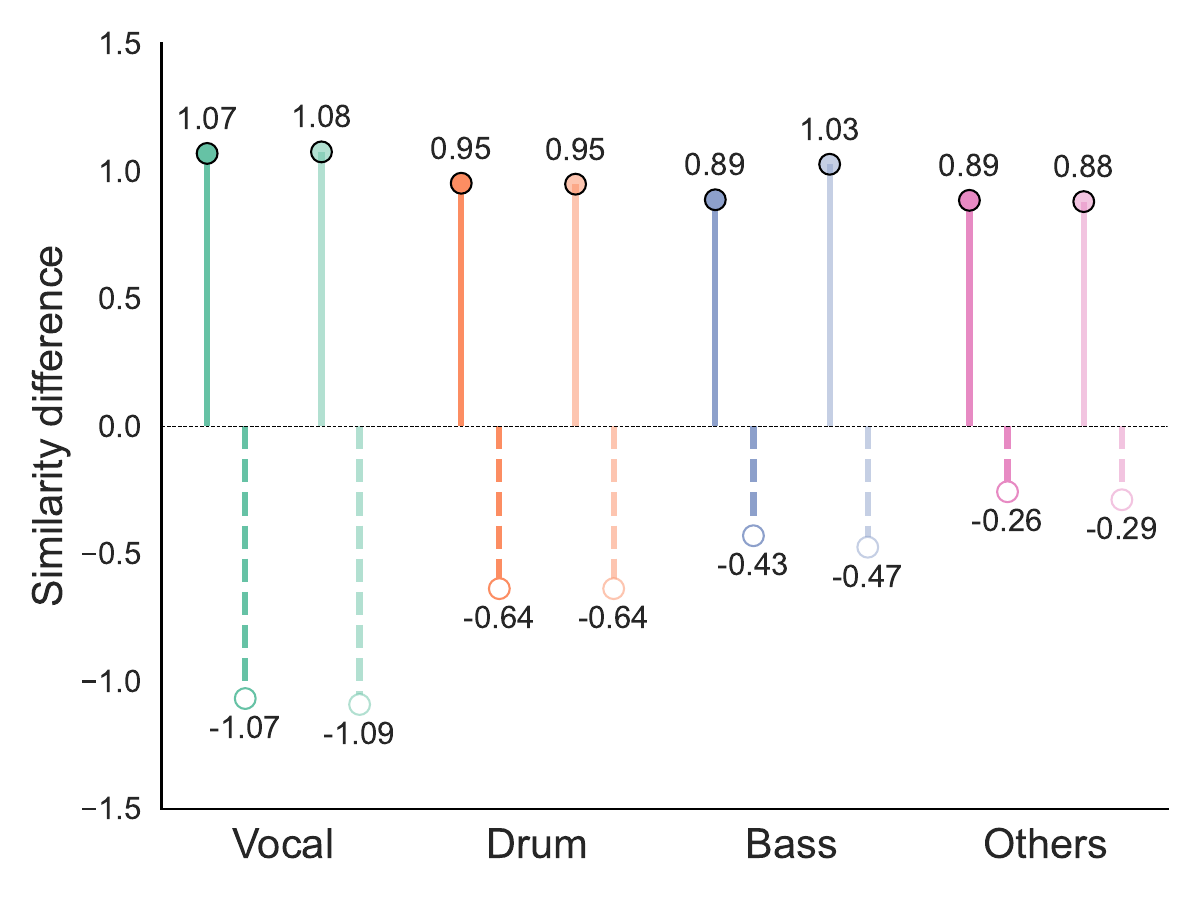}
    \caption{All-0 ms}
    \label{fig:subfig1}
\end{subfigure}
\begin{subfigure}{0.3\textwidth}
    \centering
    \includegraphics[width=\linewidth]{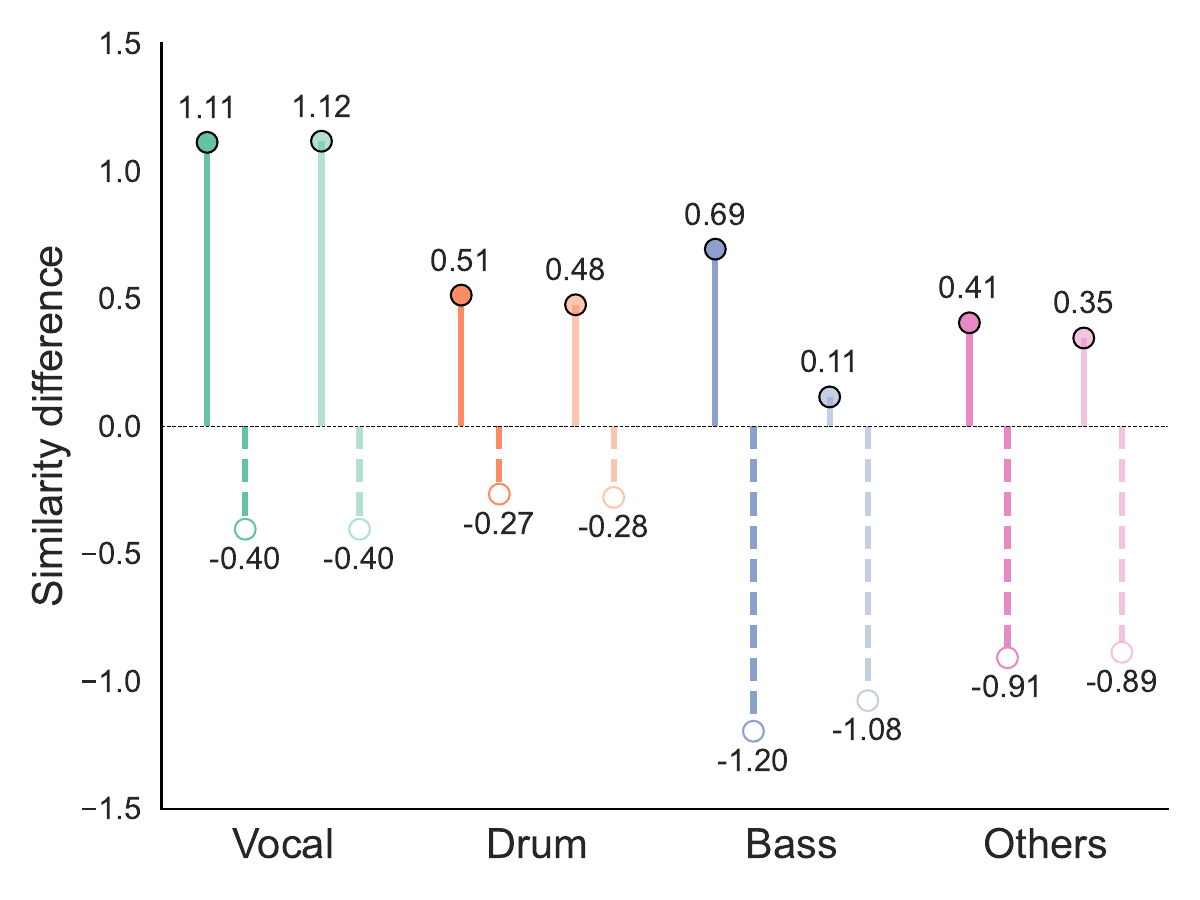}
    \caption{Attn-0 ms}
    \label{fig:subfig2}
\end{subfigure}
\begin{subfigure}{0.3\textwidth}
    \centering
    \includegraphics[width=\linewidth]{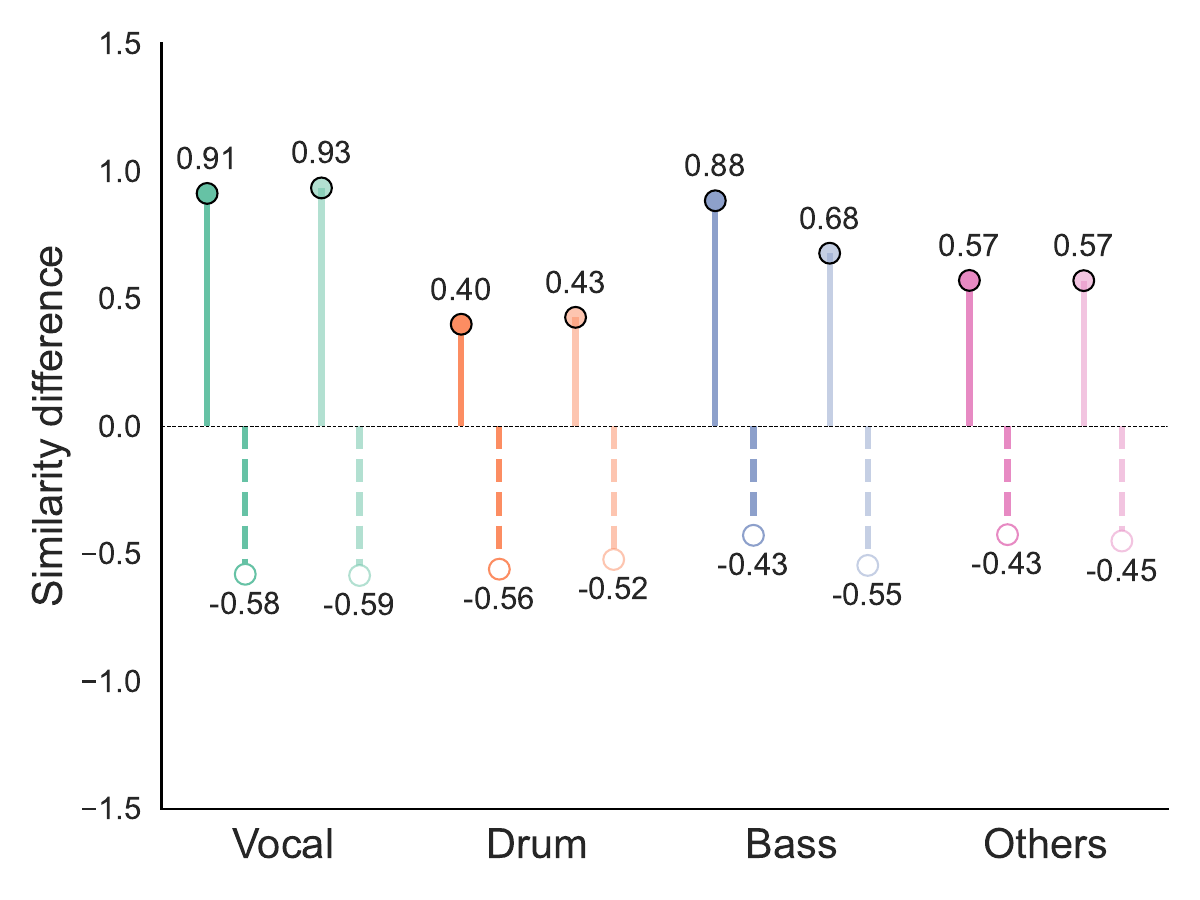}
    \caption{All-200 ms}
    \label{fig:subfig3}
\end{subfigure}

\caption{Task-level similarity difference vectors.}
\label{fig:4-diff}
\begin{minipage}{0.8\textwidth}
\normalsize
Each graph is divided into the \textbf{Correctly Matched} and \textbf{Incorrectly Matched} groups. Dark colors represent evaluation on all data, whereas lighter colors represent evaluation on high-attention data. \textbf{Model: all-0 ms} exhibited consistently stable differences across all tasks in the \textbf{Correctly Matched} group, while showing larger differences in the \textit{Vocal} task within the \textbf{Incorrectly Matched} group compared to the other two models. In contrast, \textbf{Model: attn-0 ms} demonstrated significantly larger differences in the \textbf{Incorrectly Matched} group for the \textit{Bass} and \textit{Others} tasks.
\end{minipage}
\end{figure}

We then compute a 4$\times$1 vector of task-level similarity difference. Figure \ref{fig:4-diff} presents the results, where solid lines denote the \textbf{Correctly Matched} group and dashed lines denote the \textbf{Incorrectly Matched} group. Within each group, dark colors represent evaluation on all data, whereas lighter colors represent evaluation on high-attention data.

In the \textbf{Correctly Matched} group, \textbf{Model: all-0 ms} demonstrated consistently stable similarity differences across all tasks. Notably, for the \textit{Drum} task, \textbf{Model: all-0 ms} produced larger similarity differences than the other two models. In contrast, for the \textit{Bass} and \textit{Others} tasks, \textbf{Model: attn-0 ms} exhibited smaller similarity differences, particularly in the high-attention data evaluation. 
For the \textbf{Incorrectly Matched} group, \textbf{Model: all-0 ms} showed larger similarity differences in the \textit{Vocal} task compared to the other two models. For the \textit{Bass} and \textit{Others} tasks, \textbf{Model: attn-0 ms} yielded the largest differences. This pattern aligns with the accuracy results, as \textbf{Model: attn-0 ms} showed very low accuracy on these two tasks. The limited number of \textit{Bass} and \textit{Others} trials in the high-attention dataset suggests that the model was unable to sufficiently learn the relevant features required to accurately classify these tasks.

\subsubsection*{Pair-level evaluation}
We finally computed a 4$\times$4 matrix of pair-level accuracy to evaluate across the four tasks (\textit{vocal}, \textit{drum}, \textit{bass}, and \textit{others}). Note that the global and task-level accuracy is not simply the average of the 4$\times$4 matrix of pair-level accuracy (1-vs-1 comparisons). As a result, the 4$\times$4 matrix of pair-level accuracy is not directly comparable to the global or task-level accuracy.

\begin{figure}[H]
\centering
\begin{subfigure}{0.3\textwidth}
    \centering
    \includegraphics[width=\linewidth]{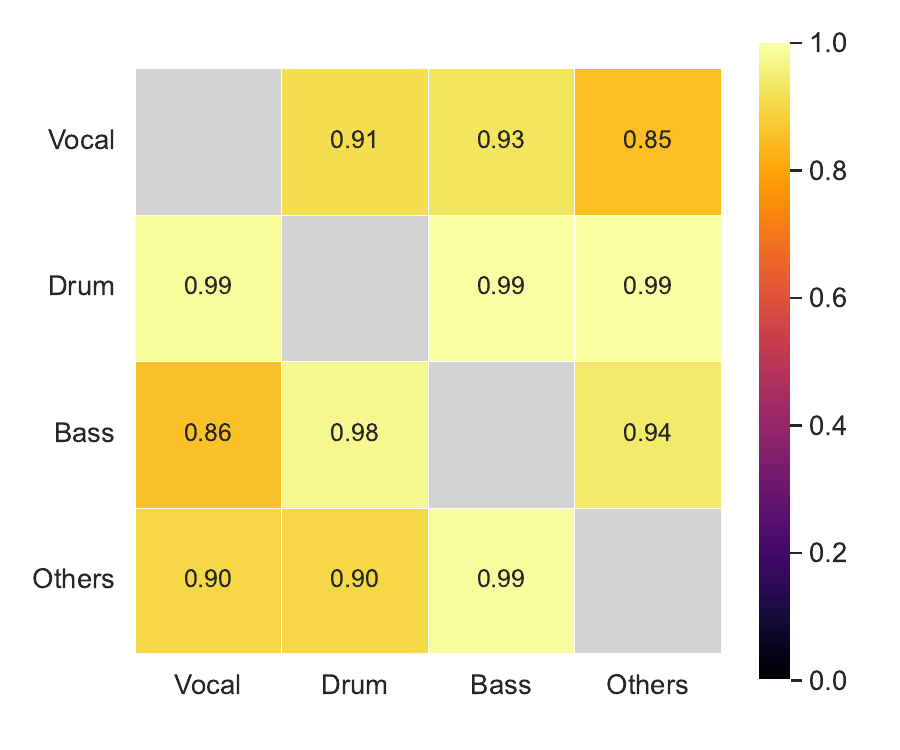}
    \caption{All-0 ms (all-data)}
    \label{fig:subfig1}
\end{subfigure}
\begin{subfigure}{0.3\textwidth}
    \centering
    \includegraphics[width=\linewidth]{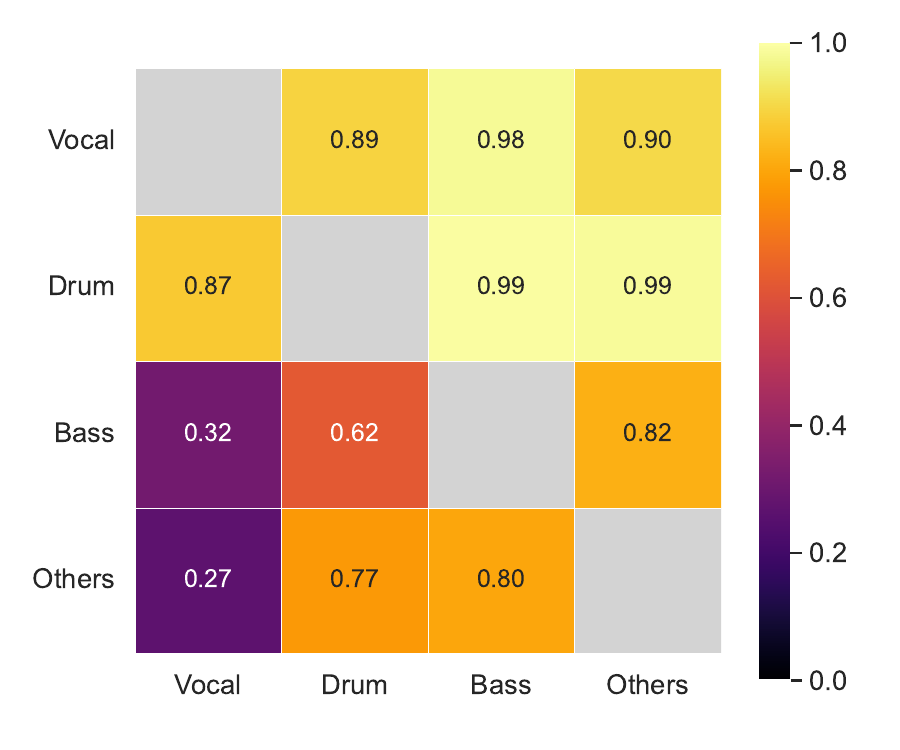}
    \caption{Attn-0 ms (all-data)}
    \label{fig:subfig2}
\end{subfigure}
\begin{subfigure}{0.3\textwidth}
    \centering
    \includegraphics[width=\linewidth]{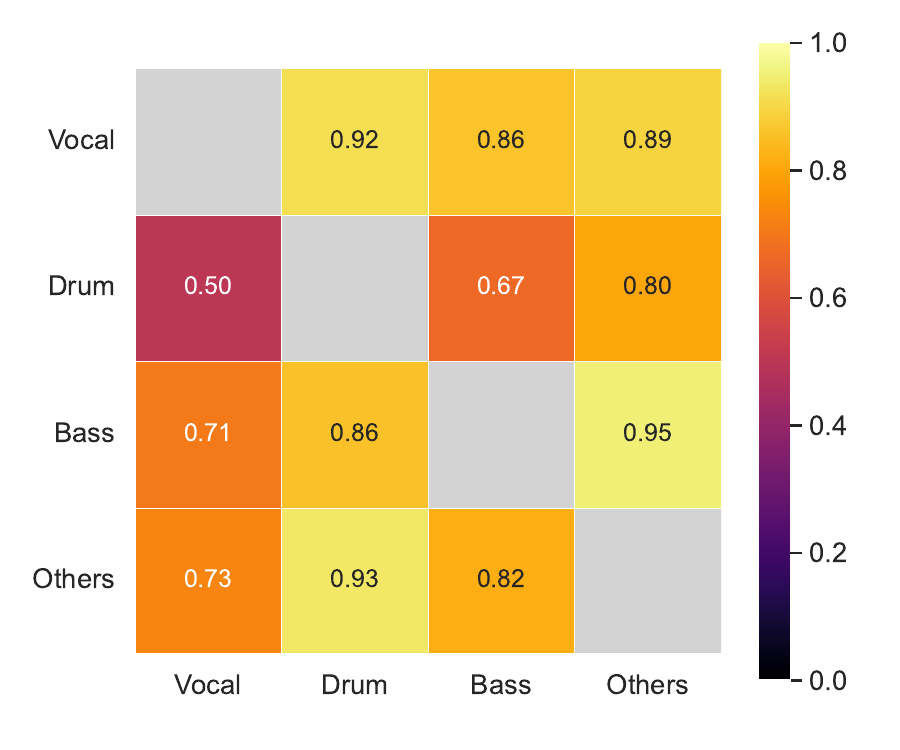}
    \caption{All-200 ms (all-data)}
    \label{fig:subfig3}
\end{subfigure}

\vspace{0.1 cm}

\begin{subfigure}{0.3\textwidth}
    \centering
    \includegraphics[width=\linewidth]{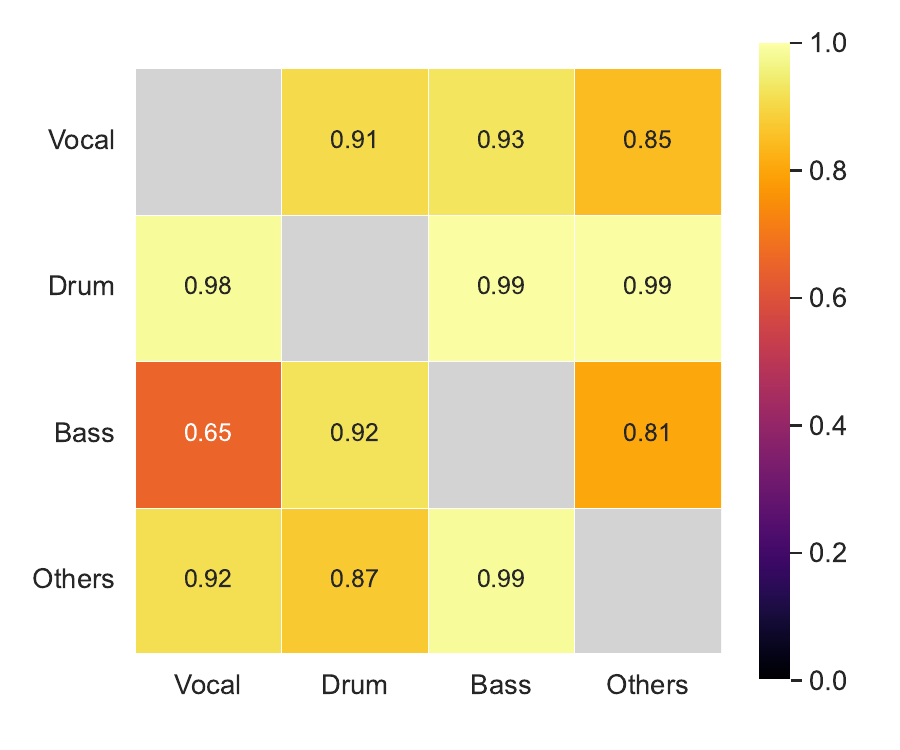}
    \caption{All-0 ms (high-attention)}
    \label{fig:subfig4}
\end{subfigure}
\begin{subfigure}{0.3\textwidth}
    \centering
    \includegraphics[width=\linewidth]{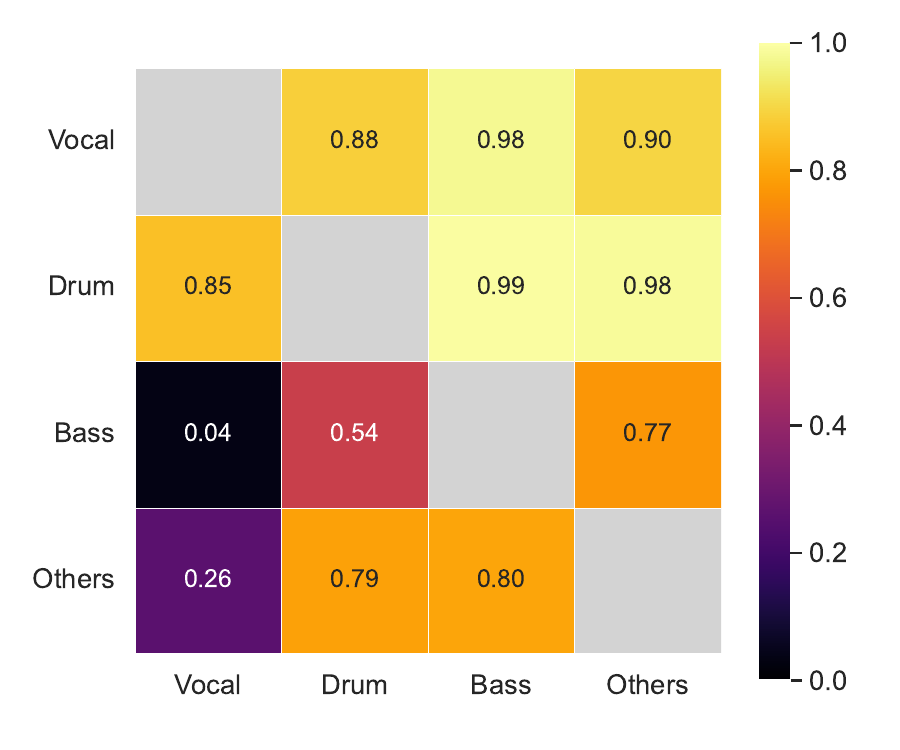}
    \caption{Attn-0 ms (high-attention)}
    \label{fig:subfig5}
\end{subfigure}
\begin{subfigure}{0.3\textwidth}
    \centering
    \includegraphics[width=\linewidth]{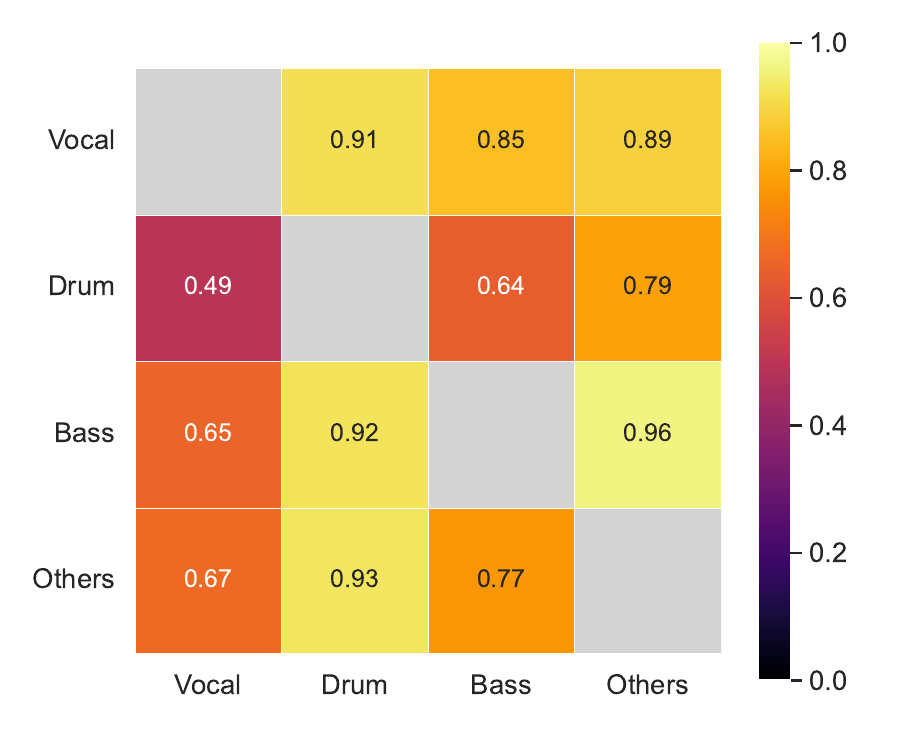}
    \caption{All-200 ms (high-attention)}
    \label{fig:subfig6}
\end{subfigure}
\caption{Accuracy matrices.}
\label{fig:matirx}
\begin{minipage}{0.8\textwidth}
\normalsize
Figure \ref{fig:matirx} (a)-(c) present the evaluation results on all test data across the three models: \textbf{Model: all-0 ms}, \textbf{Model: attn-0 ms}, and \textbf{Model: all-200 ms}. Figure \ref{fig:matirx} (d)-(f) show the corresponding results when evaluation was restricted to high-attention trials (self-reported attention scores of 4 or 5).
\textbf{Model: all-0 ms} showed relatively higher and stable accuracy across all task pairs, whereas \textbf{Model: attn-0 ms} and \textbf{Model: all-200 ms} exhibited reduced performance in specific task pairs. 
\end{minipage}
\end{figure}

Figure \ref{fig:matirx} shows the results where each entry indicates the proportion of cases in which the similarity score between EEG and audio for a positive pair (same task) exceeded that for a negative pair (different task). 

Across models, high classification accuracy was consistently observed for \textbf{Model: all-0 ms} in all-data evaluation and a drop in \textit{Bass--Vocal} and \textit{Bass--Others} in high-attention evaluation. In contrast, For the \textbf{Model: attn-0 ms}, \textit{Bass} tasks showed a pronounced drop in accuracy especially in high-attention data evaluation, particularly against \textit{Vocal} and \textit{Drum}, also the \textit{Others--Vocal} showed a very low accuracy. For the \textbf{Model: all-200 ms} tasks, \textit{Drum} showed the lowest accuracy among the three models. Also, \textit{Bass--Vocal} and \textit{Others--Vocal}, \textit{Others--Bass} showed a accuracy decline compared with \textbf{Model: all-0 ms}. This trend became more pronounced when the analysis was restricted to high-attention data, with larger gaps observed between those pairs relative to \textbf{Model: all-0 ms}, except that the \textit{Bass--Vocal} showed same accuracy (65\%). Figures \ref{fig:matirx} (d), (e), and (f) further illustrate that \textbf{Model: all-0 ms} maintained high accuracy across all task pairs except for \textit{Bass--Vocal}, while the other two models showed significant reductions in specific task pairs.

\begin{figure}[htbp]
\centering
\begin{subfigure}{0.3\textwidth}
    \centering
    \includegraphics[width=\linewidth]{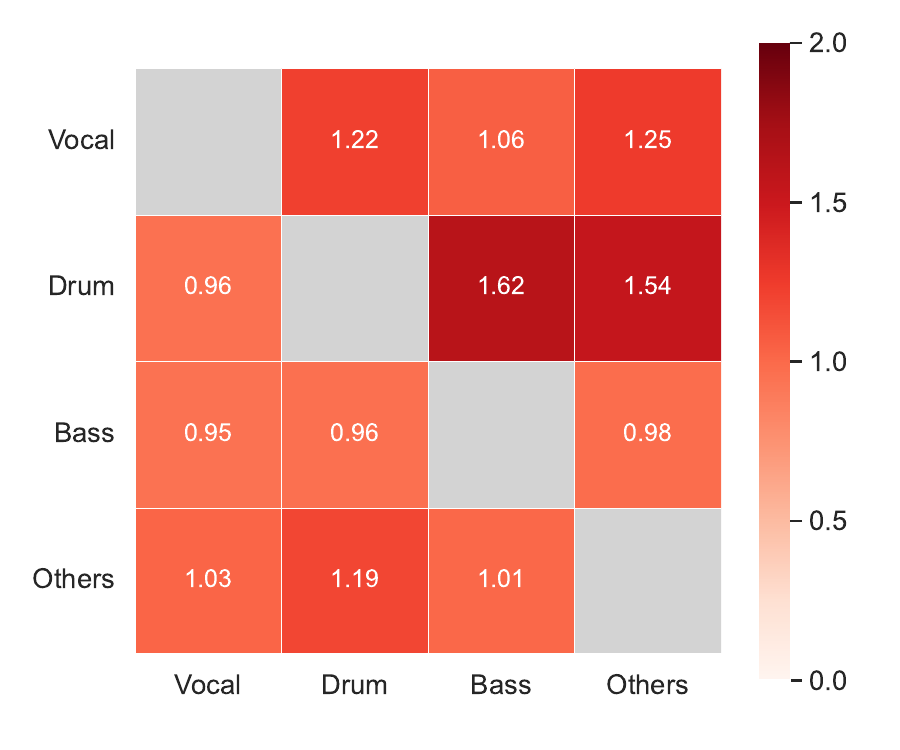}
    \caption{All-0 ms}
    \label{fig:subfig1}
\end{subfigure}
\begin{subfigure}{0.3\textwidth}
    \centering
    \includegraphics[width=\linewidth]{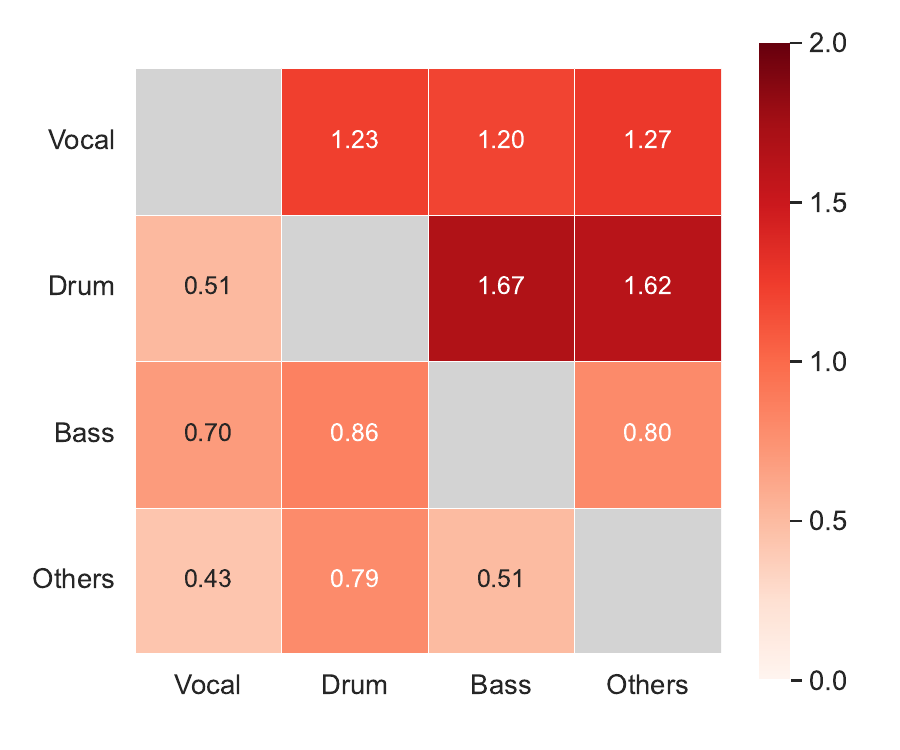}
    \caption{Attn-0 ms}
    \label{fig:subfig2}
\end{subfigure}
\begin{subfigure}{0.3\textwidth}
    \centering
    \includegraphics[width=\linewidth]{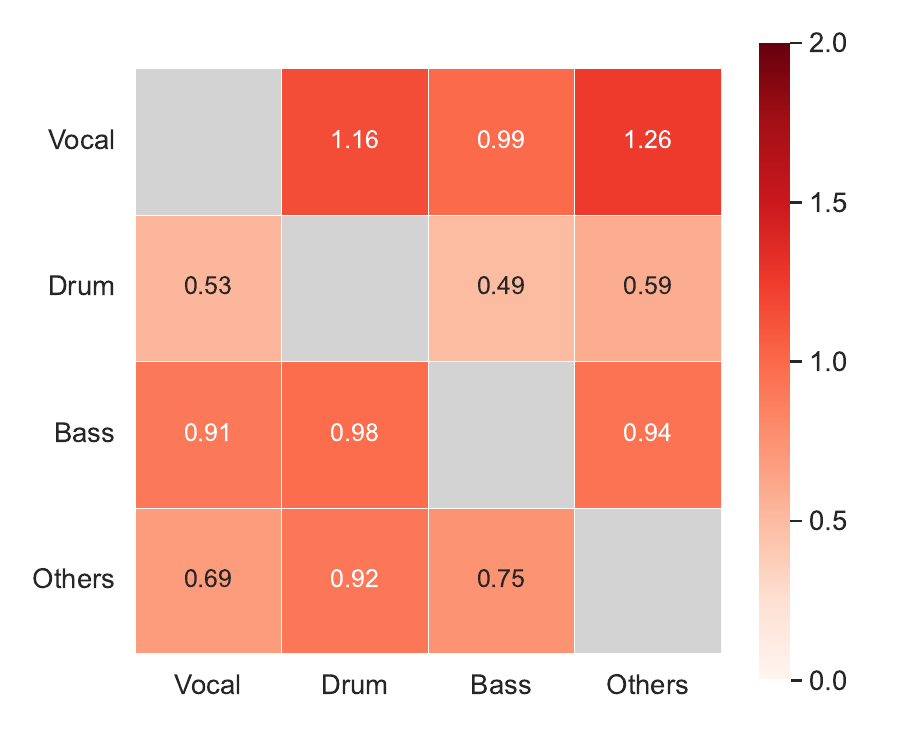}
    \caption{All-200 ms}
    \label{fig:subfig3}
\end{subfigure}

\vspace{0.1cm}

\begin{subfigure}{0.3\textwidth}
    \centering
    \includegraphics[width=\linewidth]{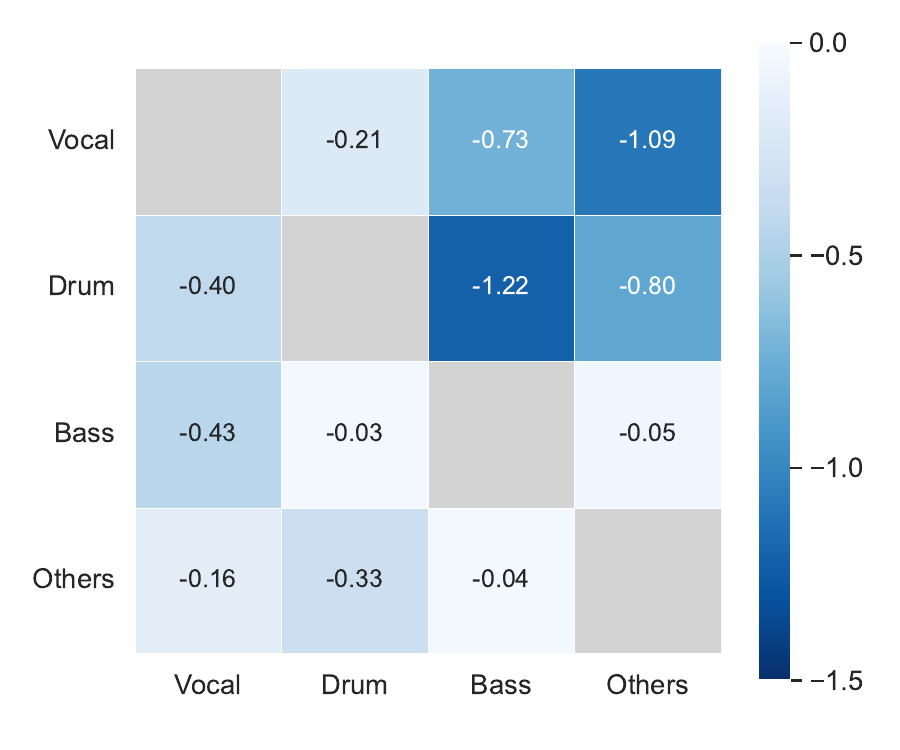}
    \caption{All-0 ms}
    \label{fig:subfig4}
\end{subfigure}
\begin{subfigure}{0.3\textwidth}
    \centering
    \includegraphics[width=\linewidth]{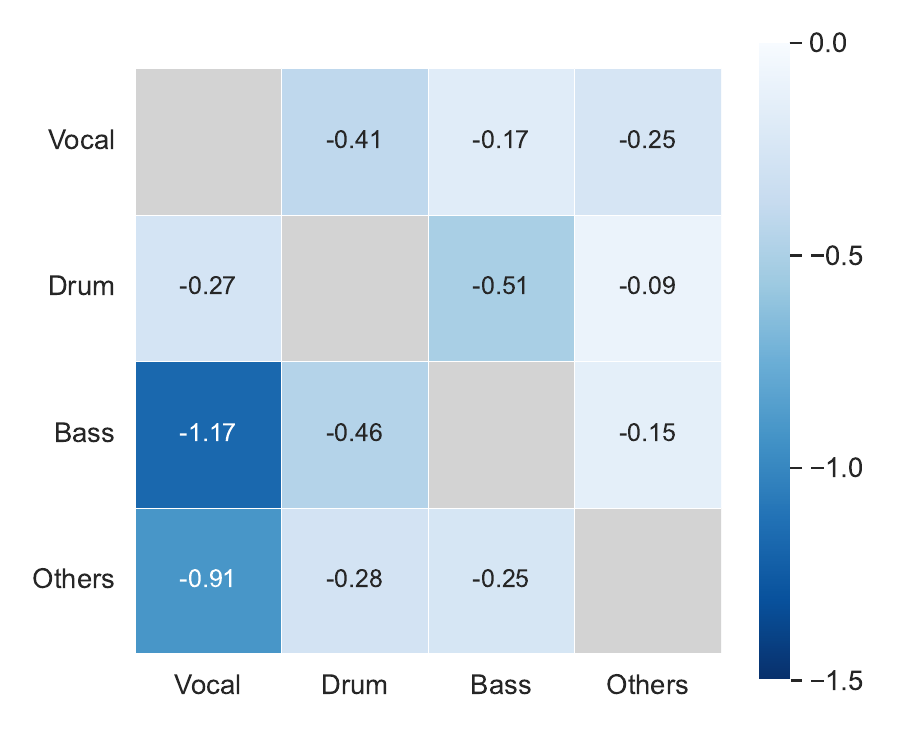}
    \caption{Attn-0 ms}
    \label{fig:subfig5}
\end{subfigure}
\begin{subfigure}{0.3\textwidth}
    \centering
    \includegraphics[width=\linewidth]{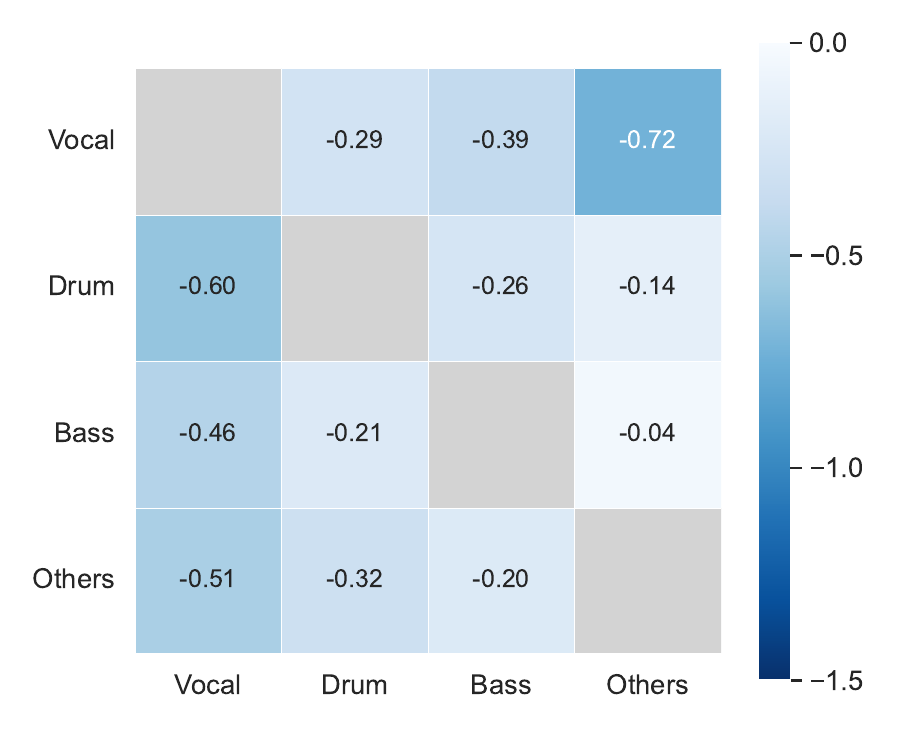}
    \caption{All-200 ms}
    \label{fig:subfig6}
\end{subfigure}
\caption{Similarity difference matrices (all-data).}
\label{fig:diff-matrix}
\begin{minipage}{0.8\textwidth}
\normalsize
Panels (a)–(c) correspond to the \textbf{Correctly Matched} group, while panels (d)–(f) correspond to the \textbf{Incorrectly Matched} group. \textbf{Model: all-0 ms} exhibited large similarity differences in the \textbf{Correctly Matched} across all pairs, indicating the stability and robustness of the model.
\end{minipage}
\end{figure}

\begin{figure}[htbp]
\centering
\begin{subfigure}{0.3\textwidth}
    \centering
    \includegraphics[width=\linewidth]{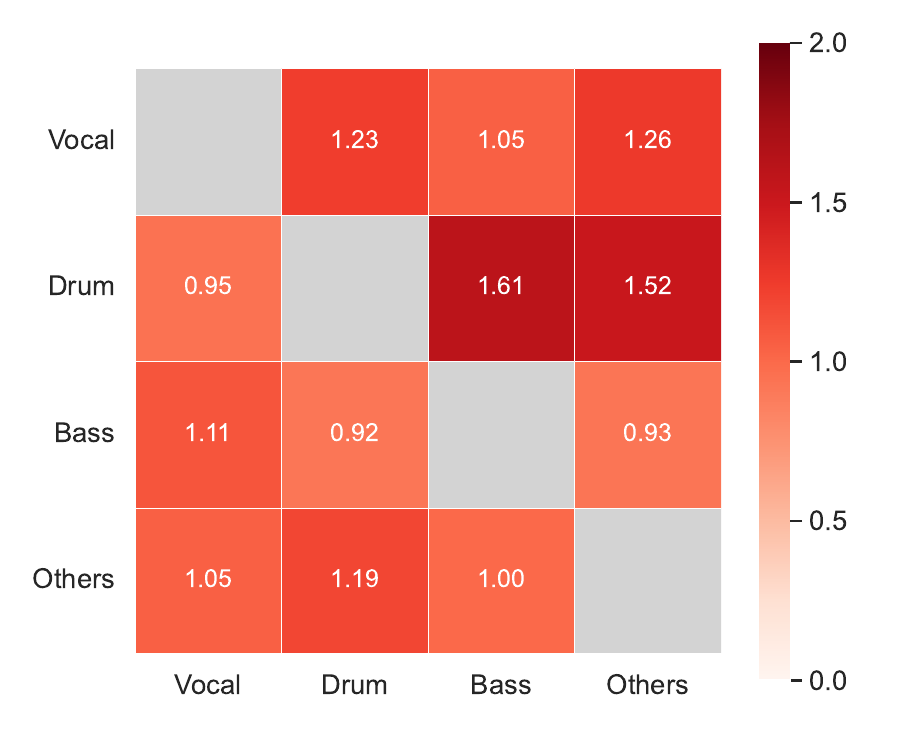}
    \caption{All-0 ms}
    \label{fig:subfig4}
\end{subfigure}
\begin{subfigure}{0.3\textwidth}
    \centering
    \includegraphics[width=\linewidth]{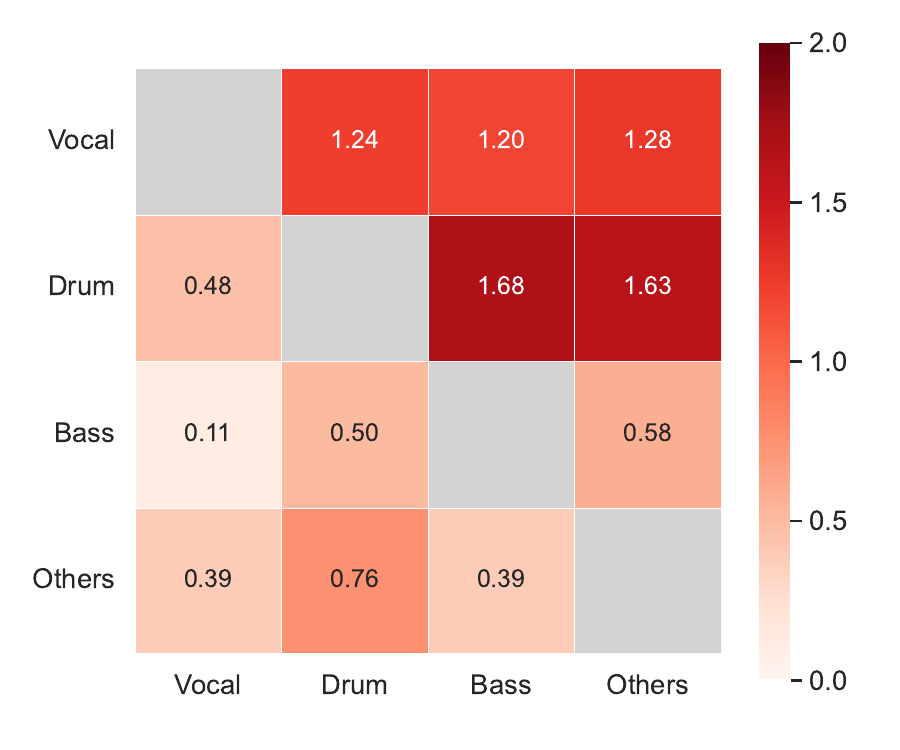}
    \caption{Attn-0 ms}
    \label{fig:subfig5}
\end{subfigure}
\begin{subfigure}{0.3\textwidth}
    \centering
    \includegraphics[width=\linewidth]{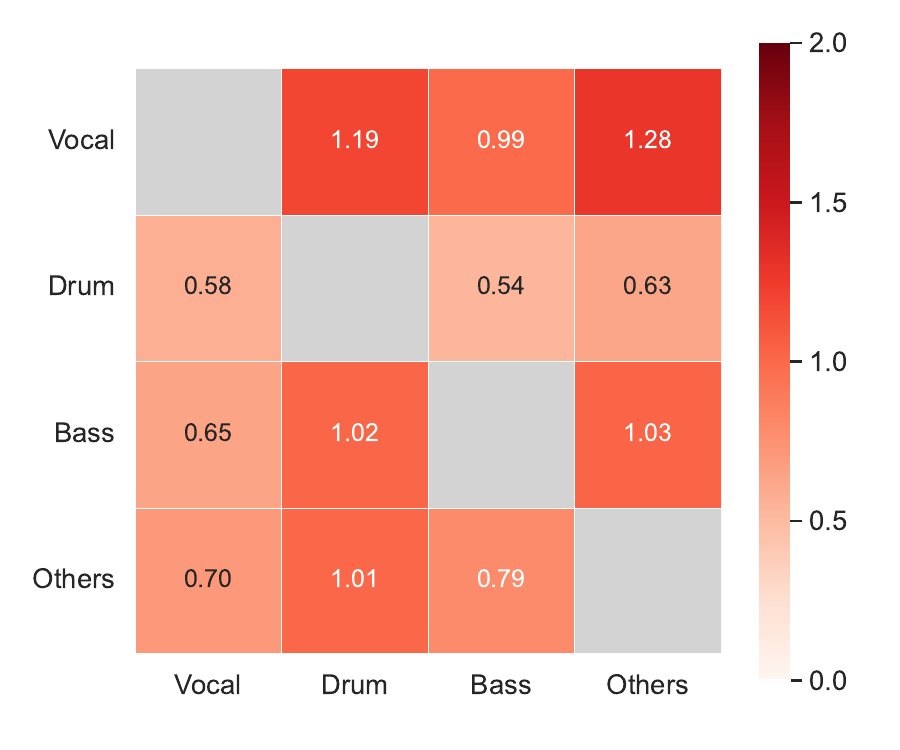}
    \caption{All-200 ms}
    \label{fig:subfig6}
\end{subfigure}

\vspace{0.1cm}

\begin{subfigure}{0.3\textwidth}
    \centering
    \includegraphics[width=\linewidth]{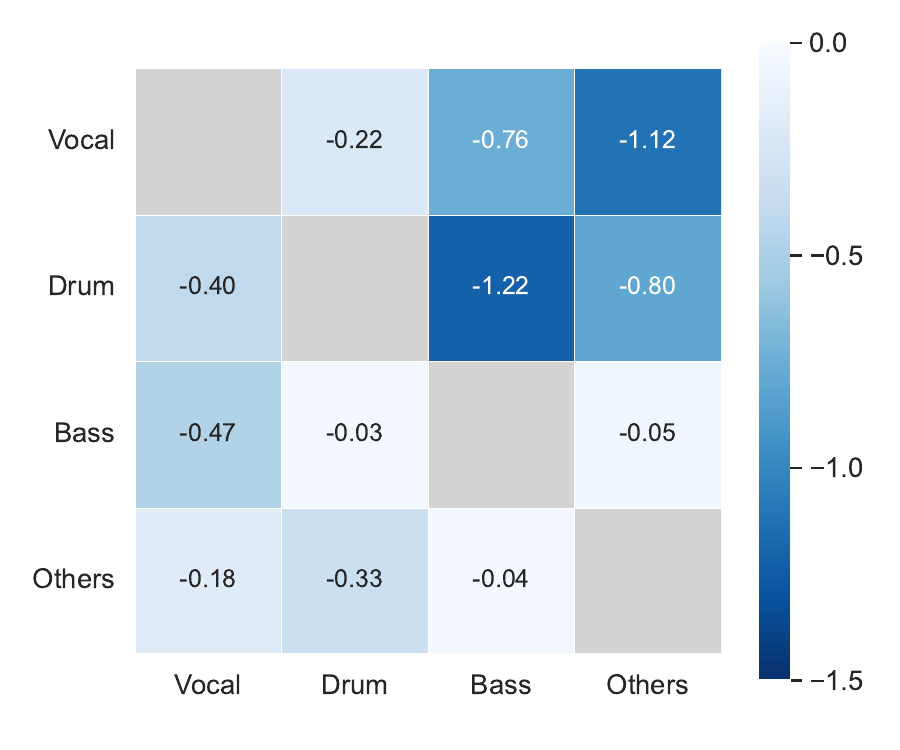}
    \caption{All-0 ms}
    \label{fig:subfig1}
\end{subfigure}
\begin{subfigure}{0.3\textwidth}
    \centering
    \includegraphics[width=\linewidth]{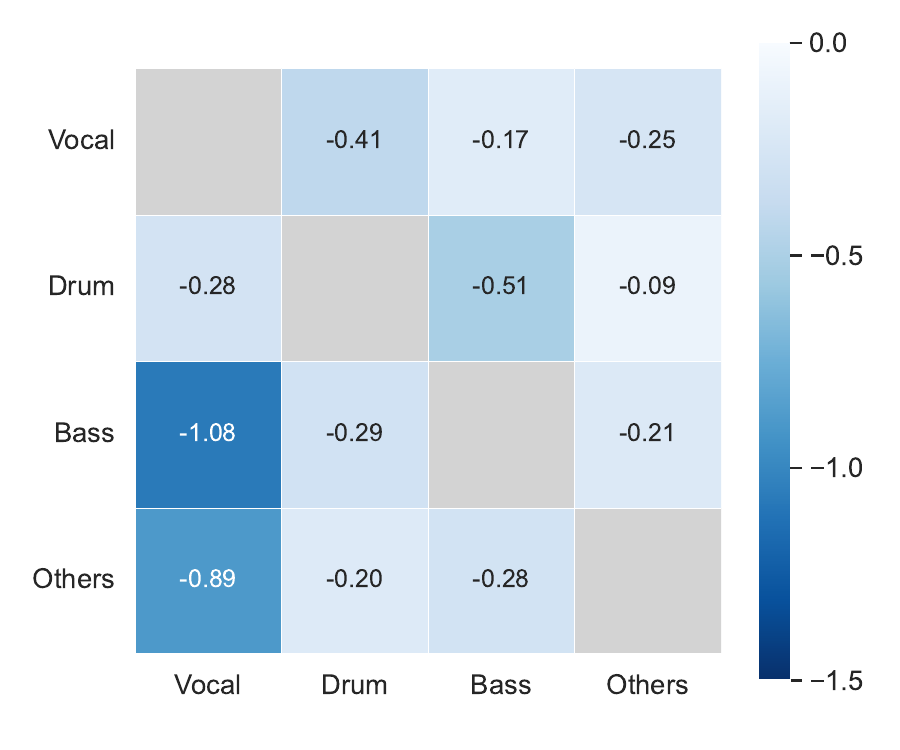}
    \caption{Attn-0 ms}
    \label{fig:subfig2}
\end{subfigure}
\begin{subfigure}{0.3\textwidth}
    \centering
    \includegraphics[width=\linewidth]{New_figs/all_200_neg_all.pdf}
    \caption{All-200 ms}
    \label{fig:subfig3}
\end{subfigure}
\caption{Similarity difference matrices (high-attention).}
\label{fig:diff-matirx-attn}
\begin{minipage}{0.8\textwidth}
\normalsize
Panels (a)–(c) correspond to the \textbf{Correctly Matched} group, while panels (d)–(f) correspond to the \textbf{Incorrectly Matched} group. Like Figure \ref{fig:diff-matrix}, the results here revealed a similar but more pronounced trend: \textbf{Model: all-0 ms} demonstrated larger similarity differences in the \textbf{Correctly Matched} group.
\end{minipage}
\end{figure}

We then analyzed similarity values using 4$\times$4 matrix of pair-level similarity difference. The results are shown in Figures \ref{fig:diff-matrix} and \ref{fig:diff-matirx-attn}, representing evaluations on all data and high-attention data respectively. In each figure, panels (a)-(c) correspond to the \textbf{Correctly Matched} group (shown in red), while panels (d)–(f) correspond to the \textbf{Incorrectly Matched} group (shown in blue).

In Figure~\ref{fig:diff-matrix} (a)–(c), all three models exhibited pronounced similarity differences for the \textit{Vocal} task pairs, indicating strong discriminability in these pairs. However, when examining other task pairs, \textbf{Model: attn-0 ms} showed a noticeable decline in performance for the \textit{Drum--Vocal}, \textit{Bass}, and \textit{Others} pairs, while \textbf{Model: all-200 ms} demonstrated a decrease in the \textit{Drum} task pairs. In contrast, \textbf{Model: all-0 ms} maintained high performance across all pairs, with differences exceeding 0.9.  

In Figure~\ref{fig:diff-matrix} (d)–(f), which depict the \textbf{Incorrectly Matched} group, \textbf{Model: all-0 ms} showed larger similarity differences for the \textit{Vocal--Bass}, \textit{Vocal--Others}, \textit{Drum--Bass} and \textit{Drum--Others} pairs, whereas other task pairs exhibited relatively smaller differences. By contrast, \textbf{Model: attn-0 ms} displayed larger differences for comparisons such as \textit{Bass--Vocal} and \textit{Others--Vocal}, while \textbf{Model: all-200 ms} produced relatively small values across all pairs.

These trends are more pronounced when considering the high-attention data evaluation, as shown in Figure \ref{fig:diff-matirx-attn}. In the \textbf{Correctly Matched} group, \textbf{Model: all-0 ms} relatively exhibited large similarity differences between positive and negative pairs across all tasks. In contrast, \textbf{Model: attn-0 ms} showed a significant drop for the \textit{Bass} and \textit{Others}, while \textbf{Model: all-200 ms} showed substantial drops for both the \textit{Drum}.

\subsubsection*{Individual evaluation}
We then examined individual performance in greater detail using \textbf{Model: all-0 ms}. Table~\ref{tab:individual} presents the global accuracy for each participant. The global accuracy ranged from 0.7755 (Sub~\#8) to 0.978 (Sub~\#3), with a mean of 0.8641 for the all-data evaluation, and from 0.7436 (Sub~\#2) to 0.978 (Sub~\#3), with a mean of 0.8454 for the high-attention evaluation. These results indicate that, despite individual variability, the model demonstrated robust generalization across subjects.

\begin{table}[htbp]
\centering
\resizebox{\textwidth}{!}{
\begin{tabular}{lcccccccccccccccccccc}
\toprule
\diagbox{Evaluation data}{Subject} & sub1 & sub2 & sub3 & sub4 & sub5 & sub6 & sub7 & sub8 & \textbf{Mean} \\ 
\midrule
All-data & 0.9135 & 0.7981 & 0.9780 & 0.9135 & 0.8462 & 0.8077 & 0.8803 & 0.7755 & \textbf{0.8641} \\
High-attention data & 0.9231 & 0.7436 & 0.9780 & 0.9231 & 0.7500 & 0.8077 & 0.8791 & 0.7582 & \textbf{0.8454} \\
\bottomrule
\end{tabular}
}
\caption{Individual performance of \textbf{Model: all-0 ms}.}
\label{tab:individual}
\centering
\begin{minipage}{0.8\textwidth}
\normalsize
This table presents the results of eight individuals. Although some individual differences were observed, the model appears to generalize across individuals and can classify new songs within each subject.
\end{minipage}
\end{table}

To provide a more detailed perspective, we further analyzed performance at the level of individual songs within each subject. Due to space limitations, we report representative results from three subjects: the highest-performer Sub \#3, a mid-level performer Sub \#7, and the lowest-performer Sub \#2. 

\begin{figure}[H]
\centering
\begin{subfigure}{0.3\textwidth}
    \centering
    \includegraphics[width=\linewidth]{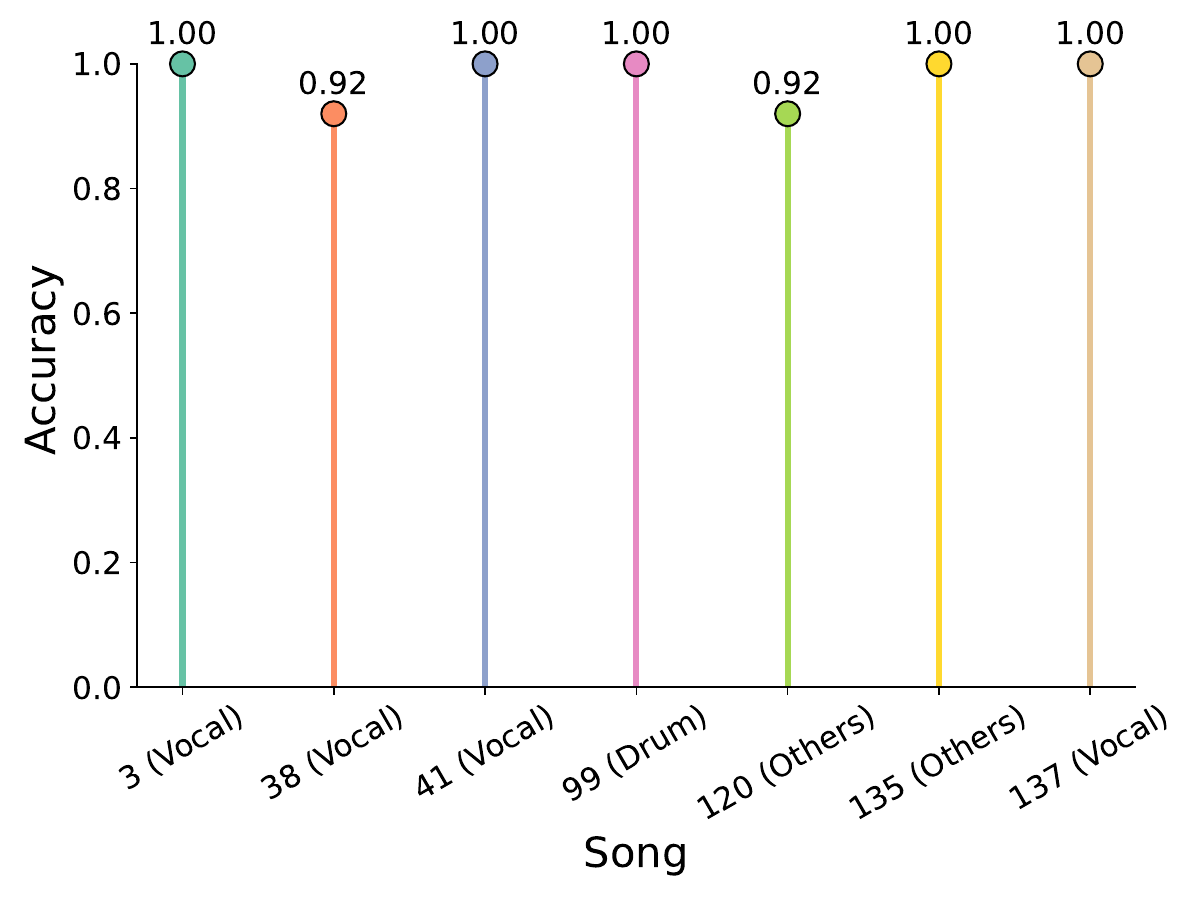}
    \caption{Sub \#3-acc}
    \label{fig:subfig1}
\end{subfigure}
\begin{subfigure}{0.3\textwidth}
    \centering
    \includegraphics[width=\linewidth]{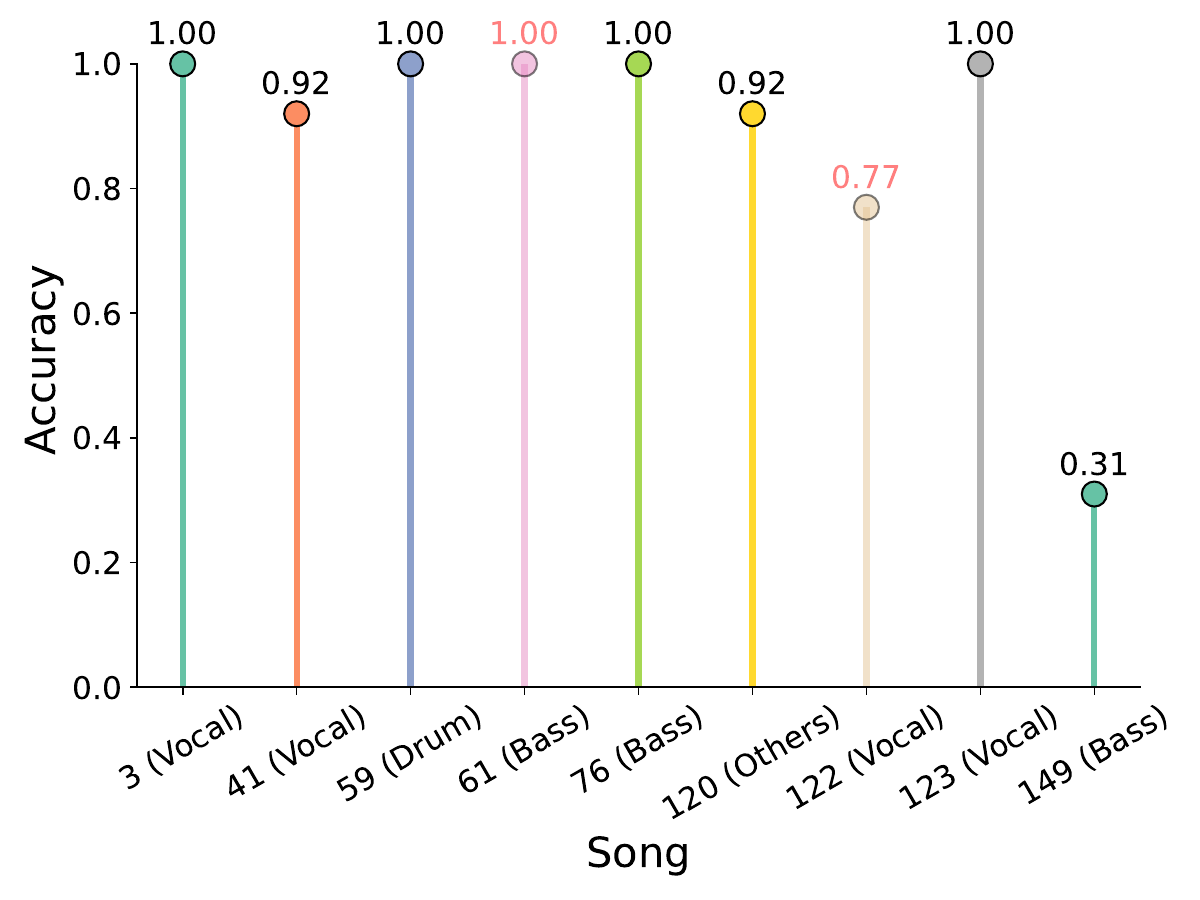}
    \caption{Sub \#7-acc}
    \label{fig:subfig2}
\end{subfigure}
\begin{subfigure}{0.3\textwidth}
    \centering
    \includegraphics[width=\linewidth]{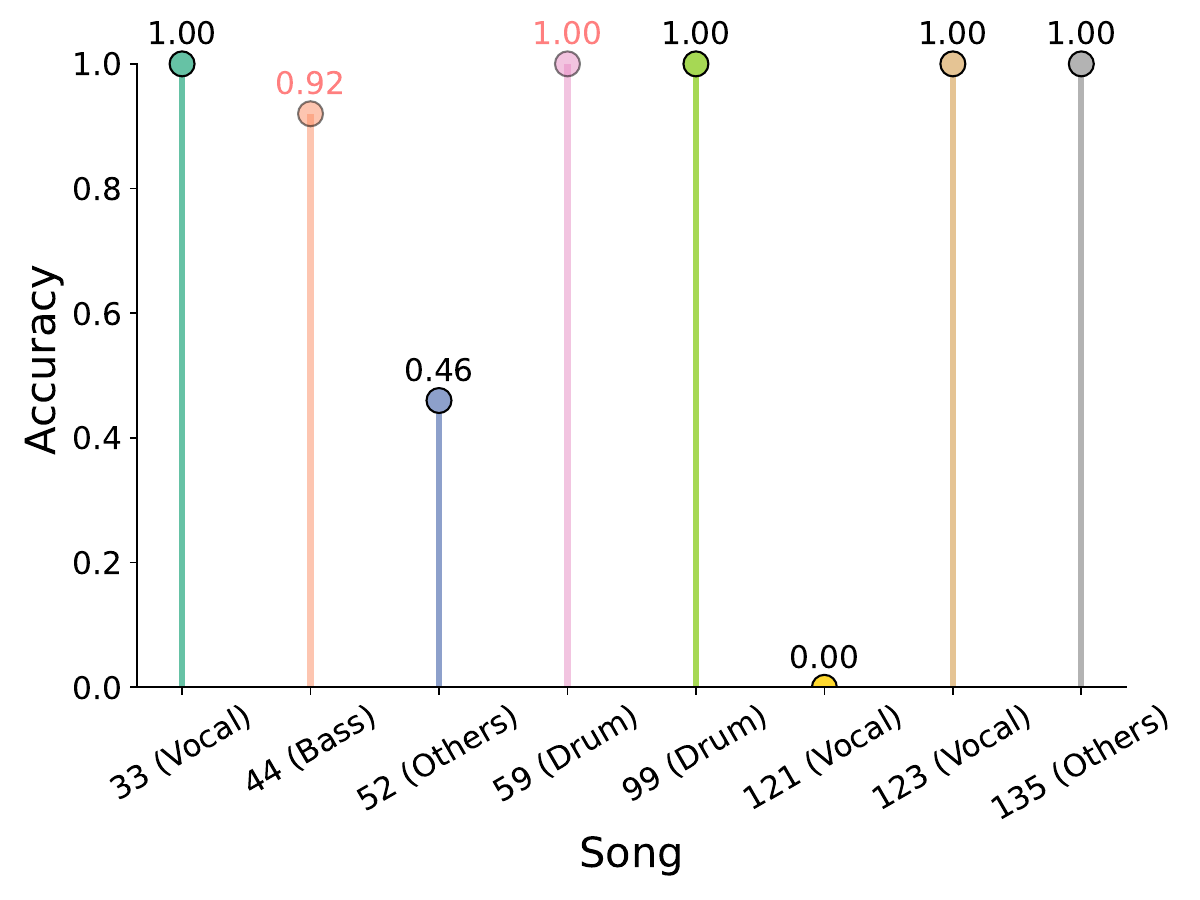}
    \caption{Sub \#2-acc}
    \label{fig:subfig3}
\end{subfigure}

\vspace{0.1cm}

\begin{subfigure}{0.3\textwidth}
    \centering
    \includegraphics[width=\linewidth]{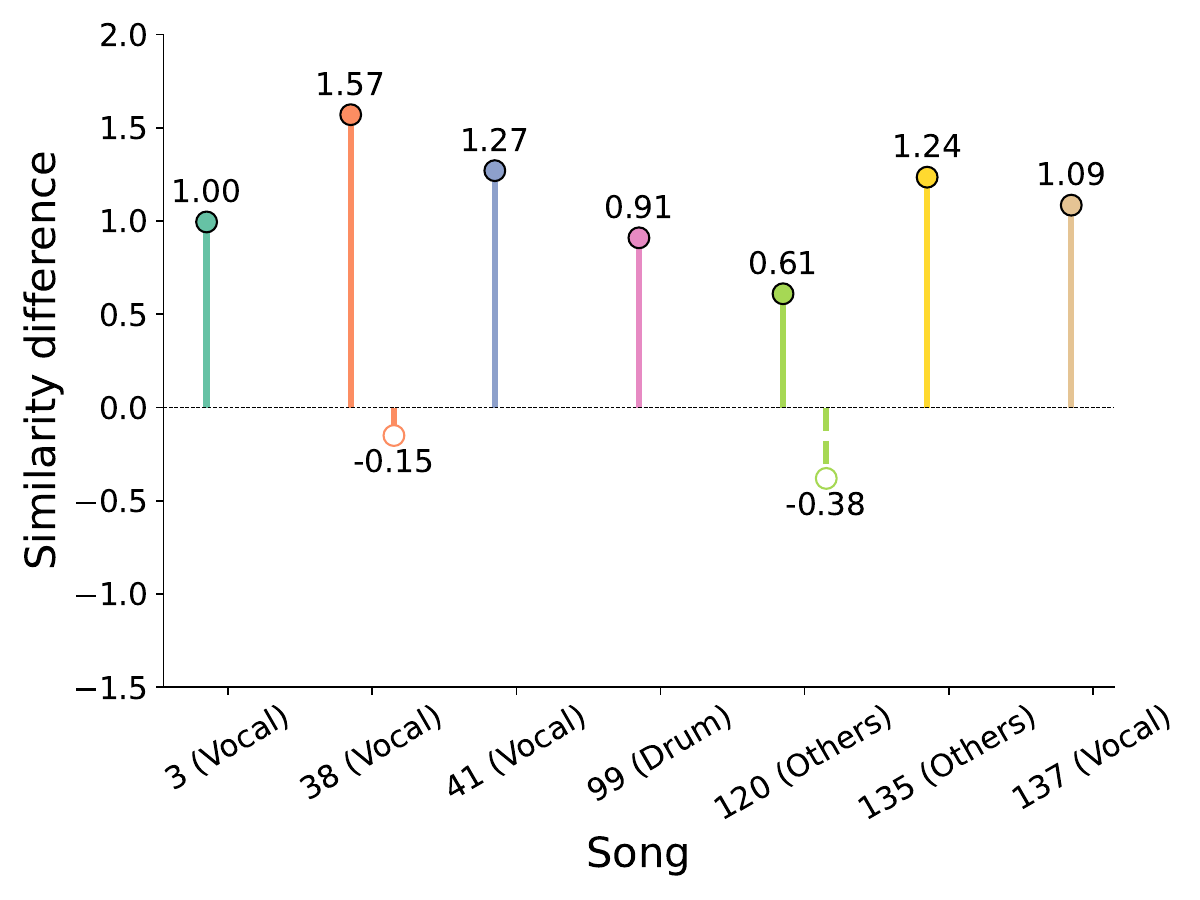}
    \caption{Sub \#3-diff}
    \label{fig:subfig4}
\end{subfigure}
\begin{subfigure}{0.3\textwidth}
    \centering
    \includegraphics[width=\linewidth]{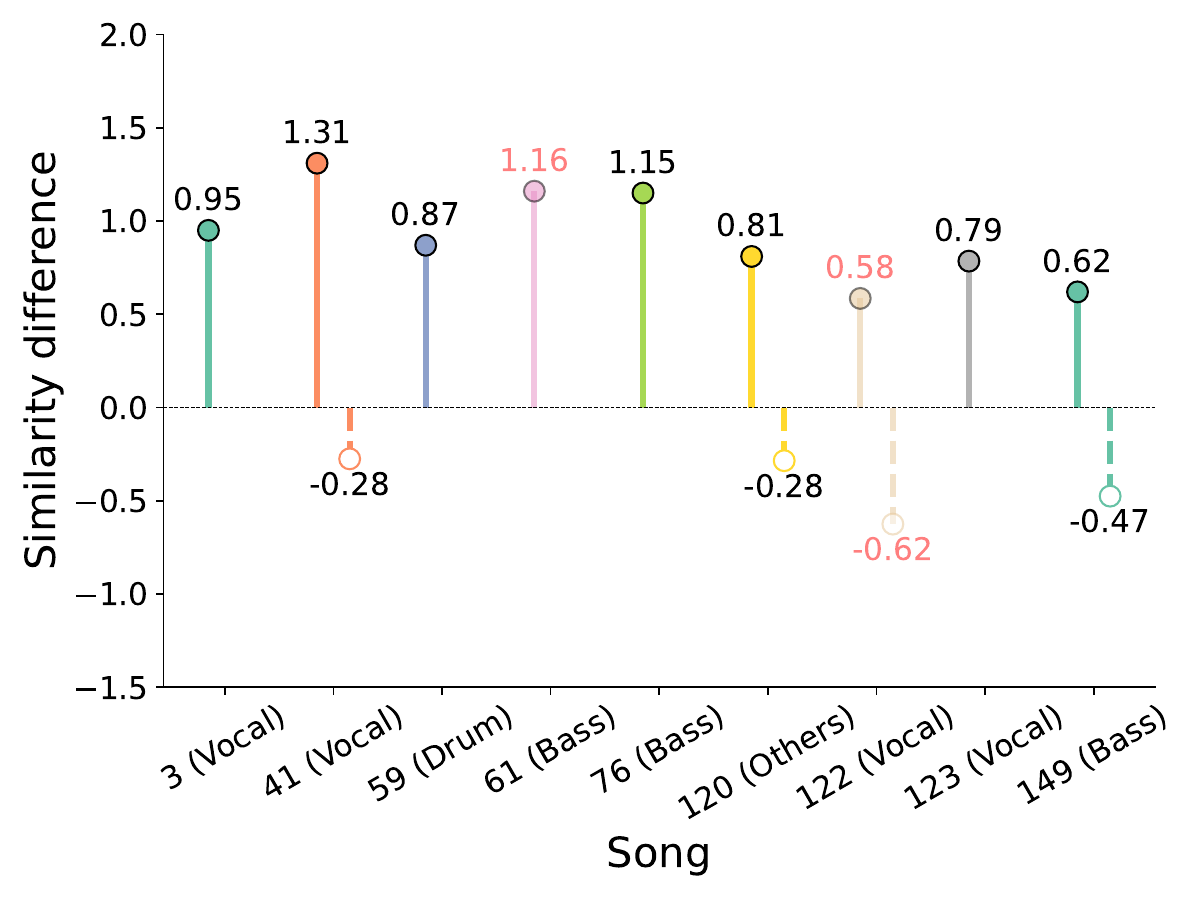}
    \caption{Sub \#7-diff}
    \label{fig:subfig5}
\end{subfigure}
\begin{subfigure}{0.3\textwidth}
    \centering
    \includegraphics[width=\linewidth]{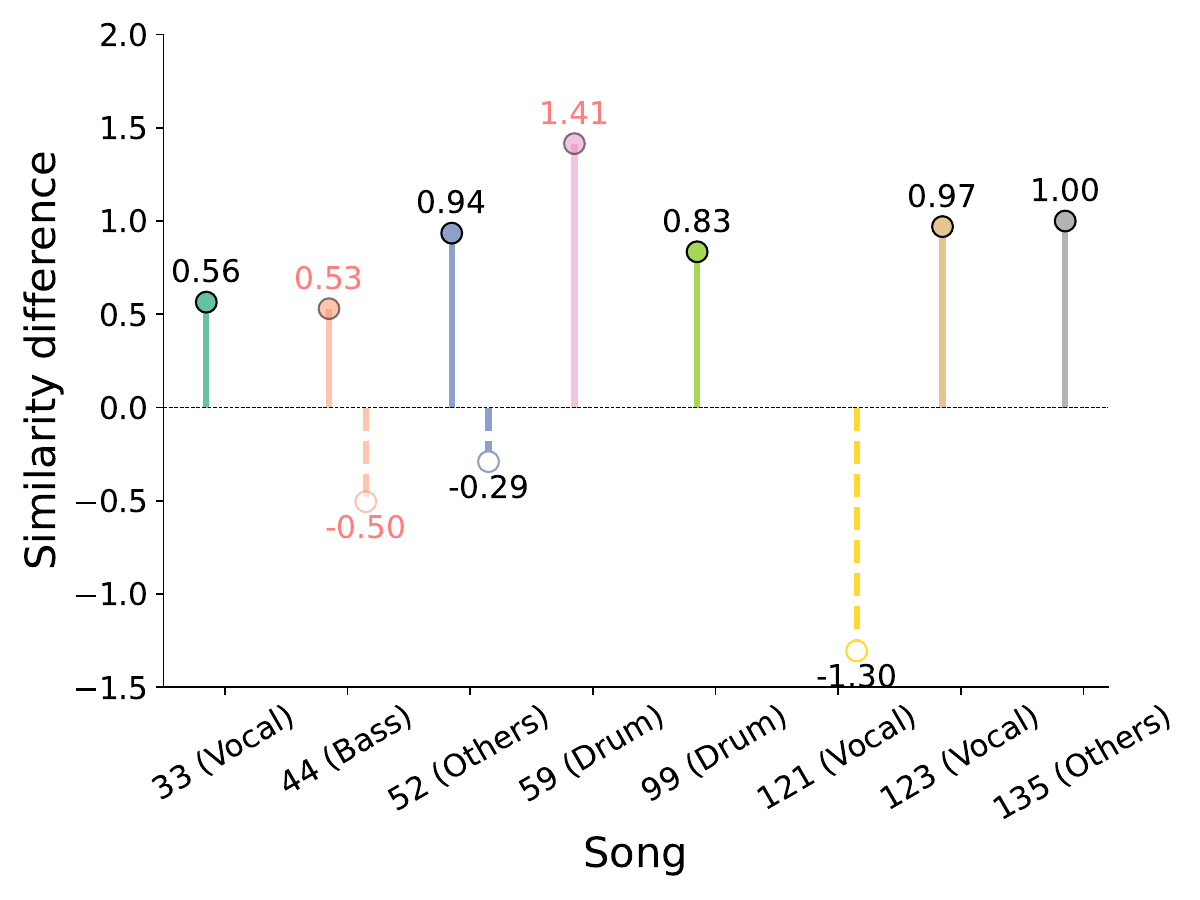}
    \caption{Sub \#2-diff}
    \label{fig:subfig6}
\end{subfigure}
\caption{Song-level performance within each subject.}
\label{fig:specific}
\begin{minipage}{0.8\textwidth}
\normalsize
Panels (a)–(c) show task-level accuracy for the three subjects, while panels (d)–(f) show task-level similarity differences. Lighter colors with red numbers indicate tasks with attention scores below 4. The best-performing subject showed consistently high performance across tasks, whereas the lowest-performing subject exhibited substantial variability across different songs.
\end{minipage}
\end{figure}

As shown in Figure \ref{fig:specific}, panels (a)–(c) present task-level accuracy results, while panels (d)–(f) show task-level similarity differences. Tasks with attention scores below 4 are shown in lighter shading, with their scores highlighted in red for clearer distinction. 
For the best-performing participant, Sub \#3, all songs achieved over 90\% accuracy, accompanied by relatively large similarity differences in the \textbf{Correctly matched} group across all tasks, and all tasks were associated with consistently high attentional engagement (attention score $>$ 4). For the intermediate performer, Sub \#7, most songs also exceeded 90\% accuracy with large differences in the \textbf{Correctly matched} group; however, certain tasks, such as \textit{song 122 (Vocal)} and \textit{song 149 (Bass)}, exhibited low accuracy and larger differences in the \textbf{Incorrectly matched} group. For the lowest-performing participant, Sub \#2, some tasks showed high accuracy, whereas others yielded very low accuracy with very large differences in the \textbf{Incorrectly matched} group. 

Considering all participants, \textit{vocal} and \textit{drum} parts generally produced high accuracy for most subjects. High performers, such as Sub \#3 and Sub \#4, showed stable performance across tasks. By contrast, other participants exhibited greater variability, performing some songs accurately while performing others poorly. This variability was particularly pronounced in low performers such as Sub \#2, whose accuracy fluctuated markedly—being relatively high for certain songs but very low for others.

Overall, the models exhibited relatively stable accuracy for participants. Among them, \textbf{Model: all-0 ms} showed relatively consistent performance across tasks, and was therefore used for subsequent analyses. 

\subsection*{Cross-subjects evaluation}
The model’s ability to generalize to new subjects is of considerable neuroscientific interest and of critical practical importance for real-world applications. We conducted cross-subject evaluations using a leave-one-subject-out scheme on the same three representative subjects as in the \textit{Individual Evaluation} section: the highest performer (Sub \#3), a mid-level performer (Sub \#7), and the lowest performer (Sub \#2).

\subsubsection*{Global evaluation}
Table \ref{tab:cross_global} reports the global accuracy and similarity difference. 
Fewer individual differences were observed across the three subjects in the similarity difference evaluation, indicating that the model is relatively robust in the cross-subject setting. The global accuracy averaged 75.56\% and 77.97\% for the two evaluations, both significantly above chance, supporting the model’s ability to generalize to new subjects.
\begin{table}[H]
\centering
\resizebox{0.7\textwidth}{!}{
\begin{tabular}{lcccccccccccccccccccc}
\toprule
\diagbox{Model}{Subject} & sub3 & sub7 & sub2 & \textbf{Mean}\\ 
\midrule
\textbf{Accuracy} (all-data)       & 0.6458 & 0.8447 & 0.7763 & \textbf{0.7556}\\
\textbf{Accuracy} (high-attention)  & 0.6862 & 0.8412 & 0.8117 & \textbf{0.7797}\\
\textbf{Difference-Correctly matched} (all-data)            & 0.87 & 0.84 & 0.85 & \textbf{0.85}\\
\textbf{Difference-Incorrectly matched} (all-data)       & -0.59 & -0.53 & -0.51 & \textbf{-0.54}\\
\textbf{Difference-Correctly matched} (high-attention)       & 0.88 & 0.84 & 0.90 & \textbf{0.87}\\
\textbf{Difference-Incorrectly matched} (high-attention)       & -0.63 & -0.55 & -0.57 & \textbf{-0.58}\\
\bottomrule
\end{tabular}
}
\caption{Global evaluation across subjects.}
\label{tab:cross_global}
\centering
\begin{minipage}{0.8\textwidth}
\normalsize
This table summarizes the results for three subjects across evaluation groups in terms of accuracy and similarity differences. The findings reveal generalizability of new subjects.
\end{minipage}
\end{table}
Notably, the subject-wise ranking was inconsistent compared with the within-subject evaluation: Sub \#7 achieved the highest accuracy, whereas Sub \#3 showed the lowest. This may be partly attributable to differences in the training datasets and to individual-specific features.

\subsubsection*{Task-level evaluation}
\begin{figure}[H]
\centering
\begin{subfigure}{0.3\textwidth}
    \centering
    \includegraphics[width=\linewidth]{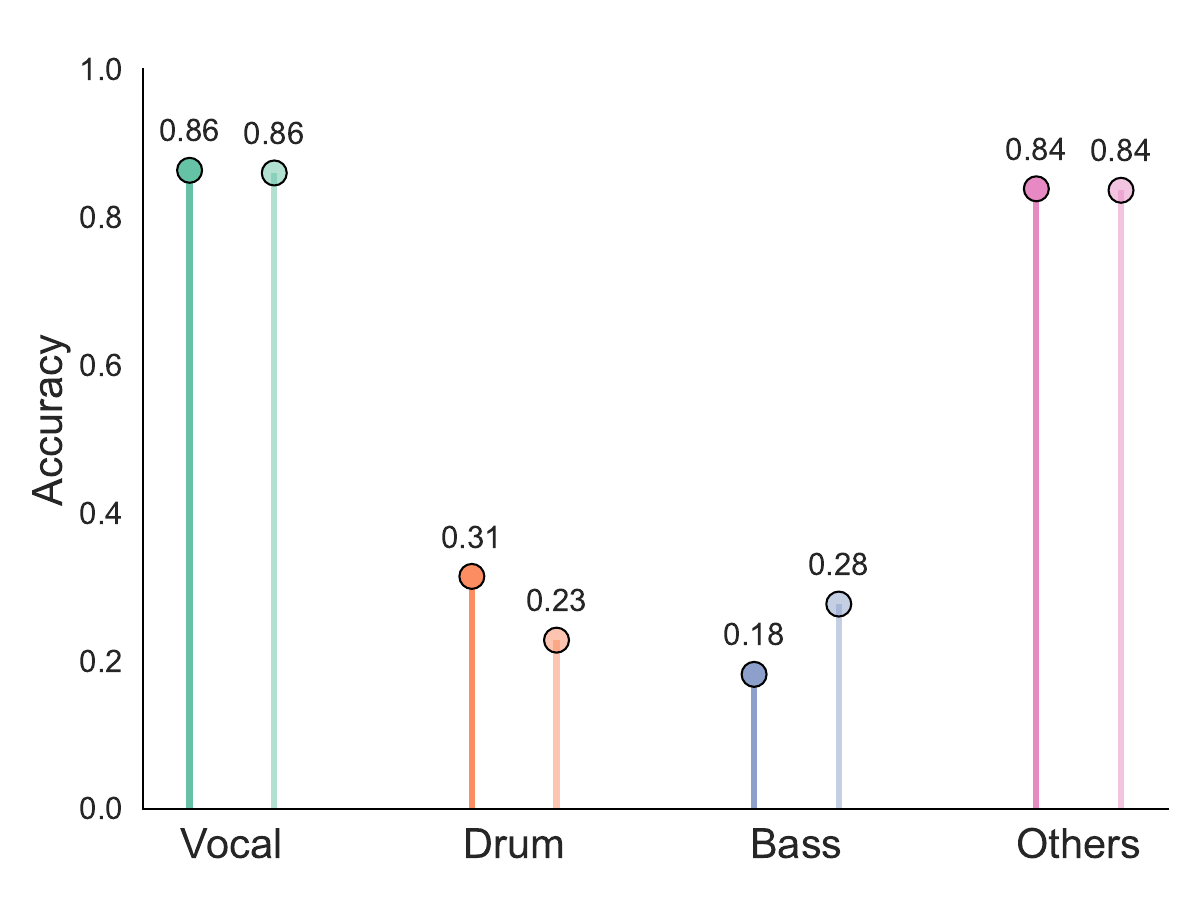}
    \caption{Sub \#3-acc}
    \label{fig:subfig1}
\end{subfigure}
\begin{subfigure}{0.3\textwidth}
    \centering
    \includegraphics[width=\linewidth]{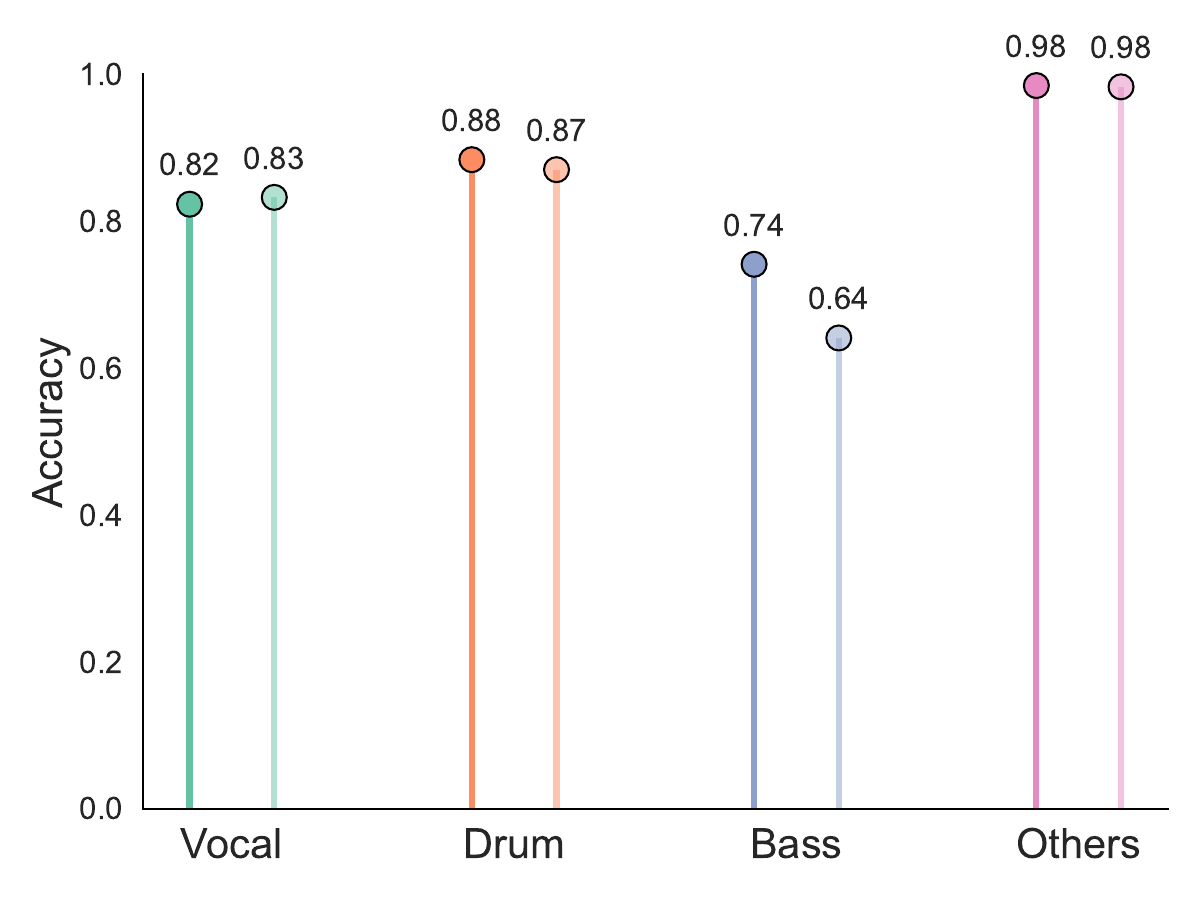}
    \caption{Sub \#7-acc}
    \label{fig:subfig2}
\end{subfigure}
\begin{subfigure}{0.3\textwidth}
    \centering
    \includegraphics[width=\linewidth]{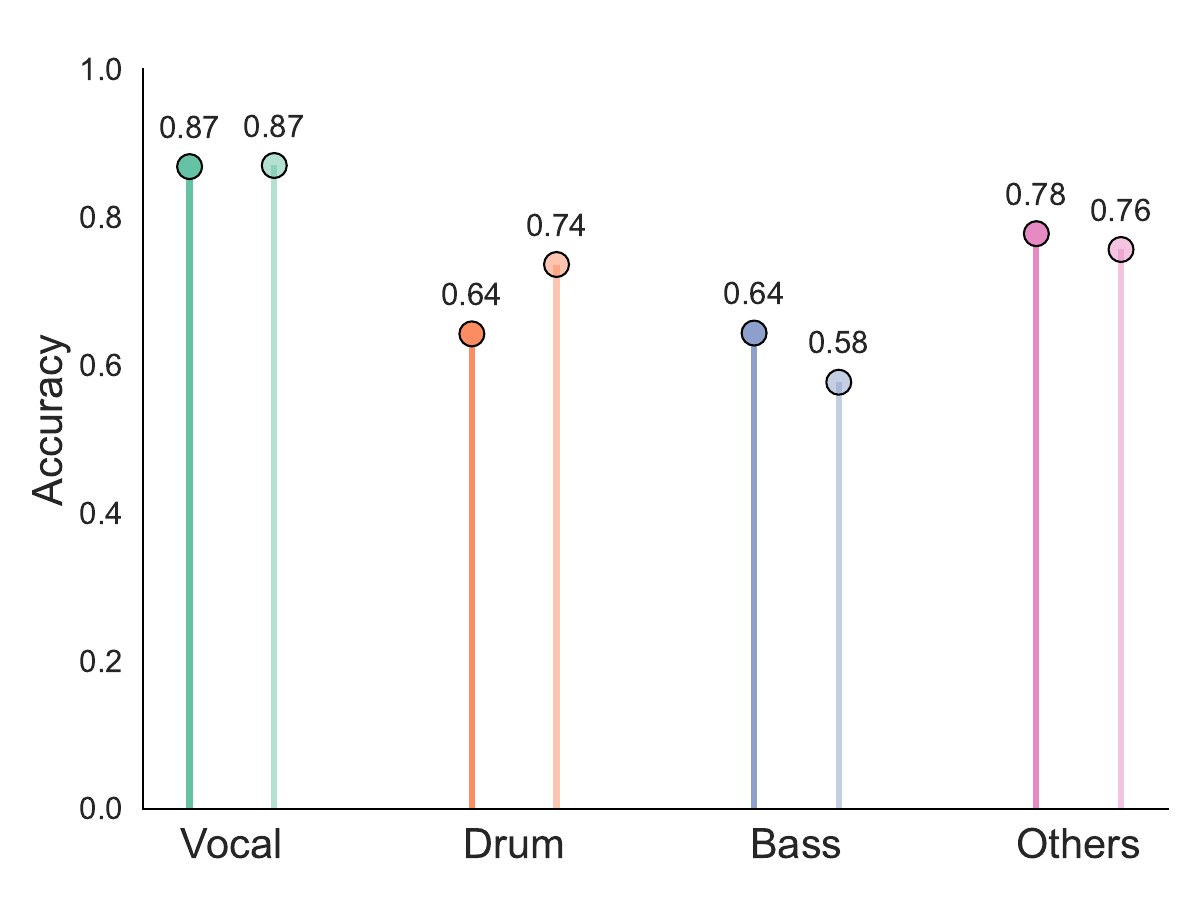}
    \caption{Sub \#2-acc}
    \label{fig:subfig3}
\end{subfigure}

\vspace{0.1cm}

\begin{subfigure}{0.3\textwidth}
    \centering
    \includegraphics[width=\linewidth]{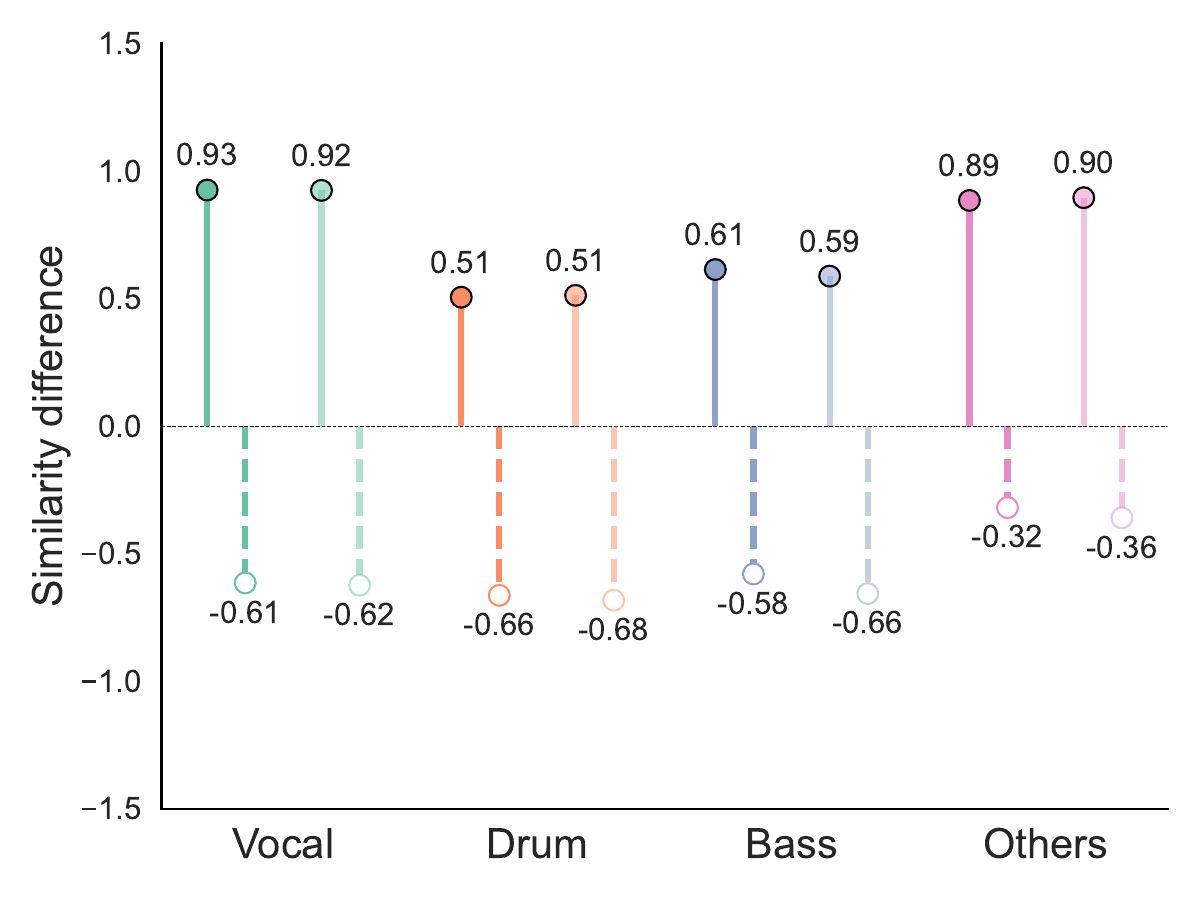}
    \caption{Sub \#3-diff}
    \label{fig:subfig4}
\end{subfigure}
\begin{subfigure}{0.3\textwidth}
    \centering
    \includegraphics[width=\linewidth]{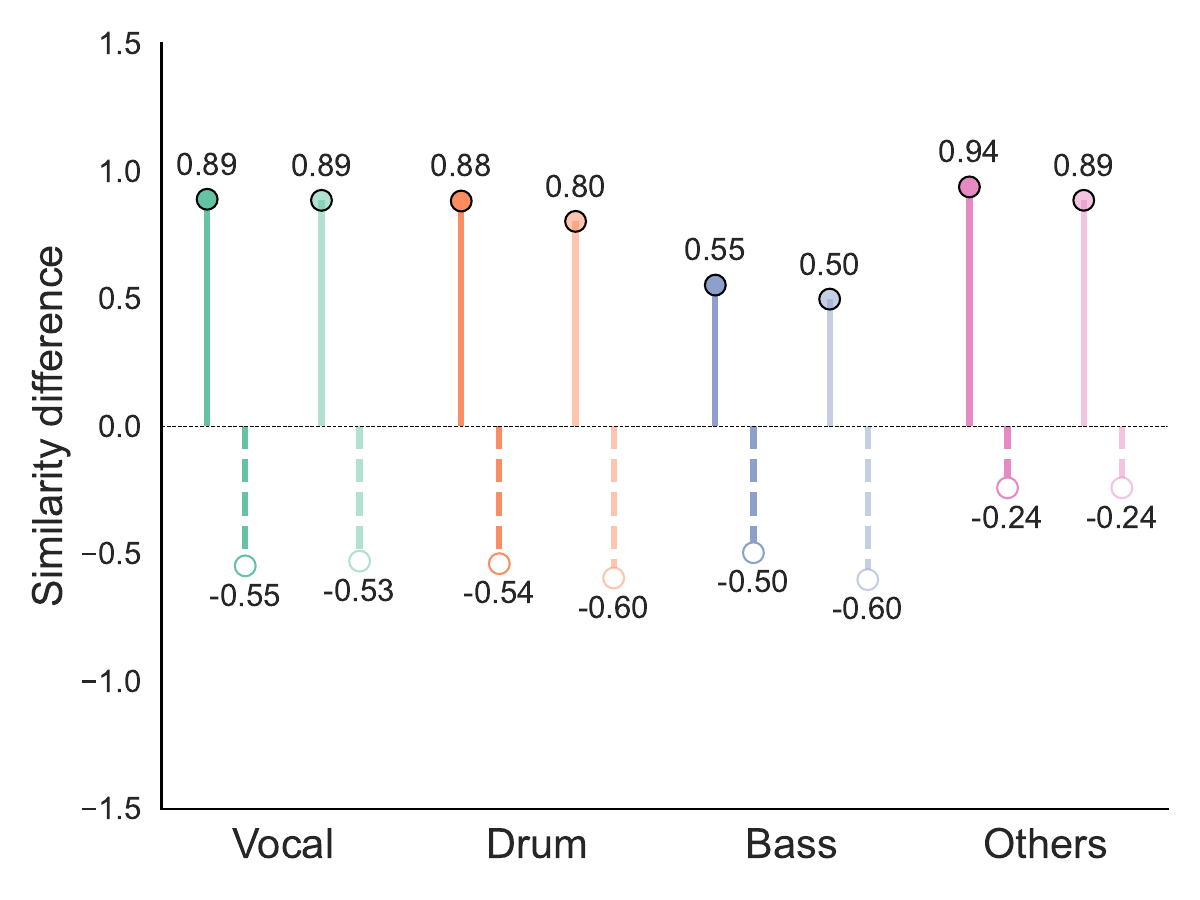}
    \caption{Sub \#7-diff}
    \label{fig:subfig5}
\end{subfigure}
\begin{subfigure}{0.3\textwidth}
    \centering
    \includegraphics[width=\linewidth]{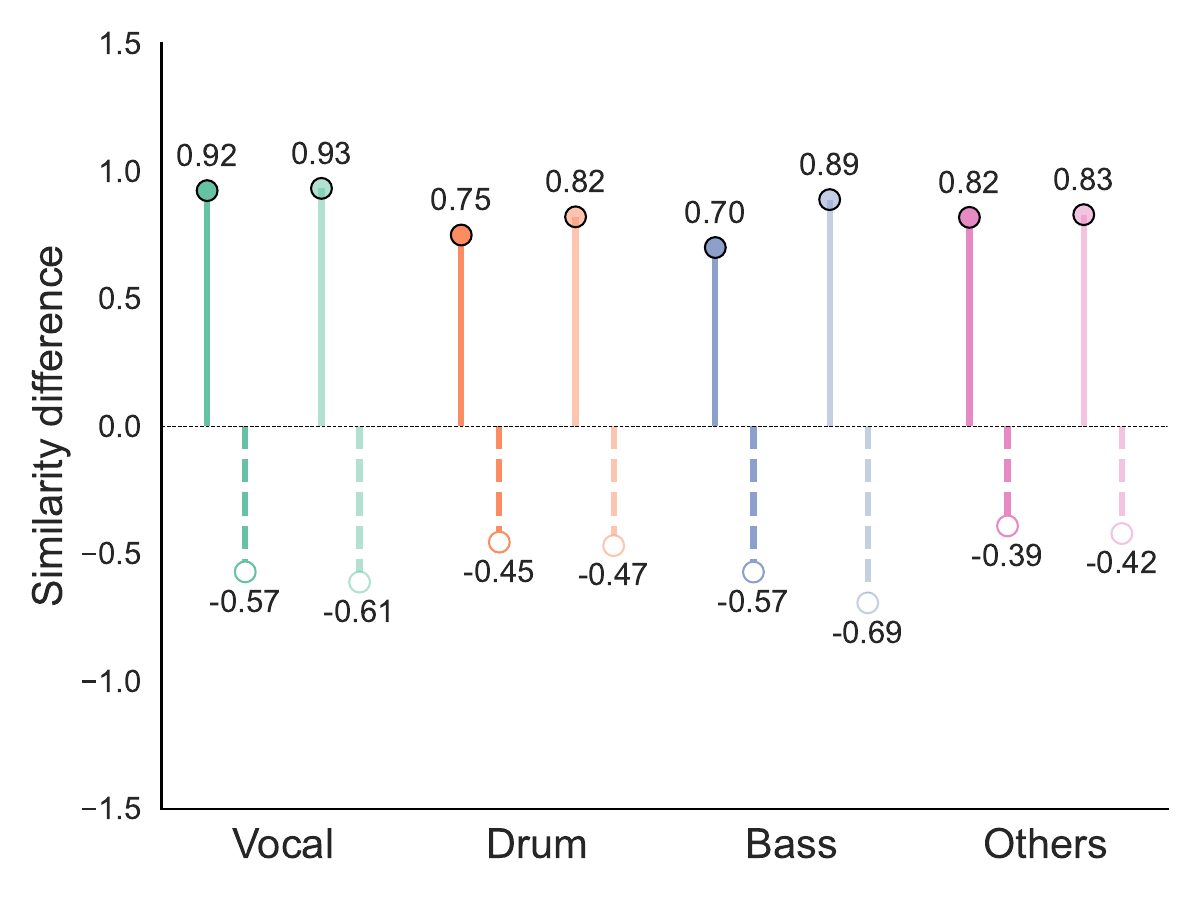}
    \caption{Sub \#2-diff}
    \label{fig:subfig6}
\end{subfigure}
\caption{Task-level evaluation across subjects.}
\label{fig:cross-sub-4}
\begin{minipage}{0.8\textwidth}
\normalsize
Panels (a)–(c) present task-level accuracy for the three subjects, while panels (d)–(f) illustrate task-level similarity differences. Lighter colors denote tasks with attention scores below 4. In contrast to the within-subject evaluation, Sub \#7 showed the best performance, whereas Sub \#3 exhibited the lowest.
\end{minipage}
\end{figure}

We then examined task-level performance (Figure \ref{fig:cross-sub-4}). Across all three subjects, the \textit{Vocal} task consistently achieved the high accuracy. For Sub\# 3, \textit{Drum} and \textit{Bass} showed relatively low performance, whereas for the other two subjects all tasks exceeded 60\% accuracy in the all-data evaluation, with Sub\# 7 exhibiting the best overall performance.
For the similarity-difference measures, fewer individual differences were observed. Sub \#3 exhibited smaller similarity differences in the \textbf{Correctly Matched} group for the \textit{Drum} and \textit{Bass} tasks, whereas Sub\# 7 showed reduced differences in the \textbf{Correctly Matched} group for \textit{Bass}.

\subsubsection*{Pair-level evaluation}
Finally, we conducted a pair-level evaluation. Due to space limitations, we focus here on the results of the accuracy matrices only.

\begin{figure}[H]
\centering
\begin{subfigure}{0.3\textwidth}
    \centering
    \includegraphics[width=\linewidth]{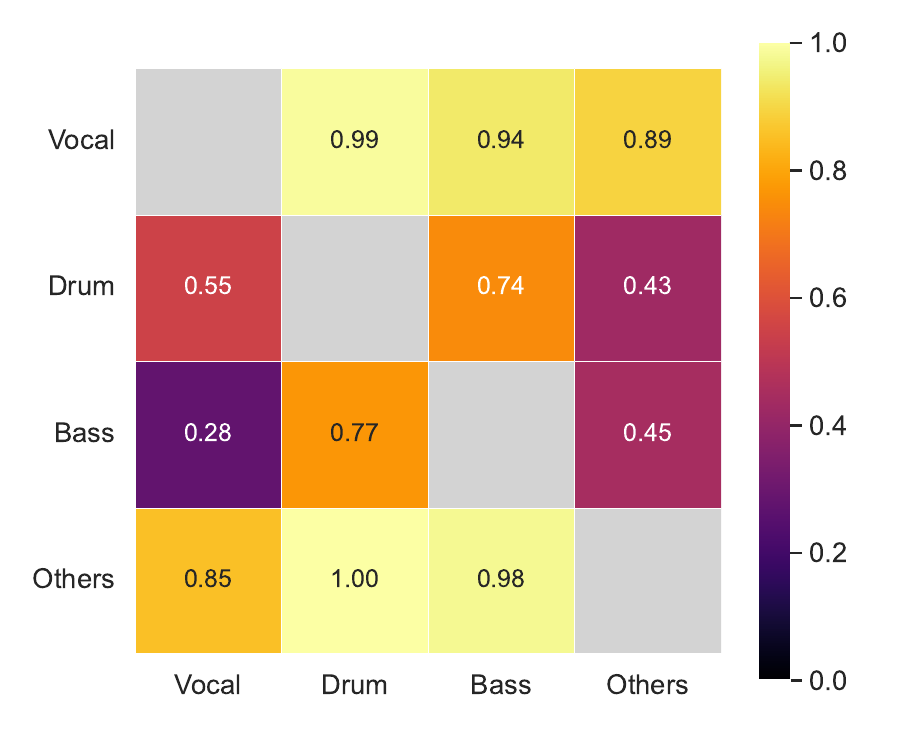}
    \caption{Sub \#3 (all-data)}
    \label{fig:subfig1}
\end{subfigure}
\begin{subfigure}{0.3\textwidth}
    \centering
    \includegraphics[width=\linewidth]{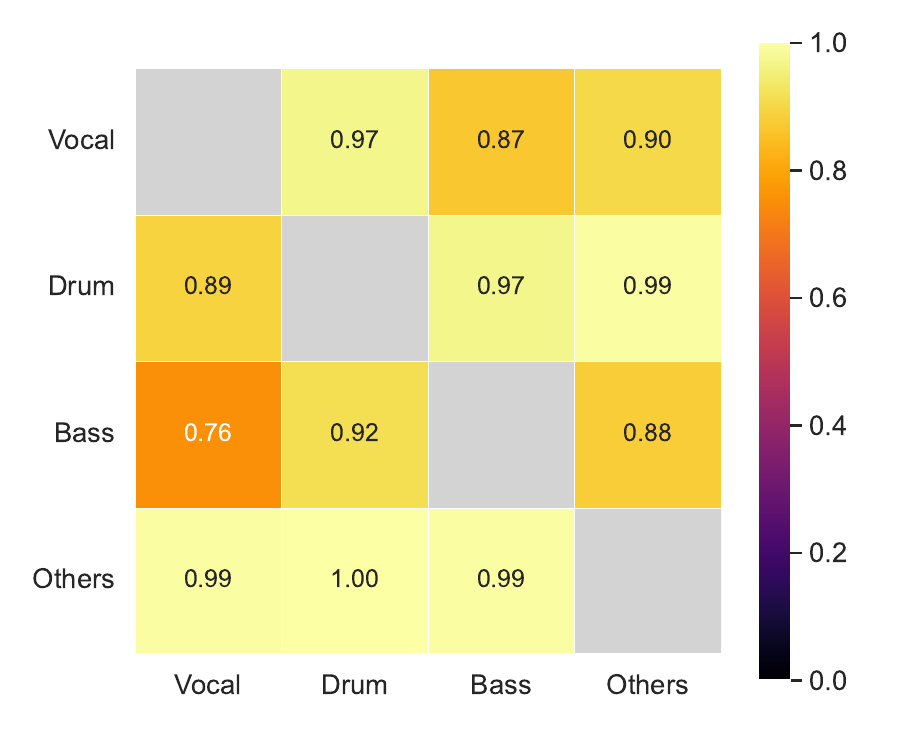}
    \caption{Sub \#7 (all-data)}
    \label{fig:subfig2}
\end{subfigure}
\begin{subfigure}{0.3\textwidth}
    \centering
    \includegraphics[width=\linewidth]{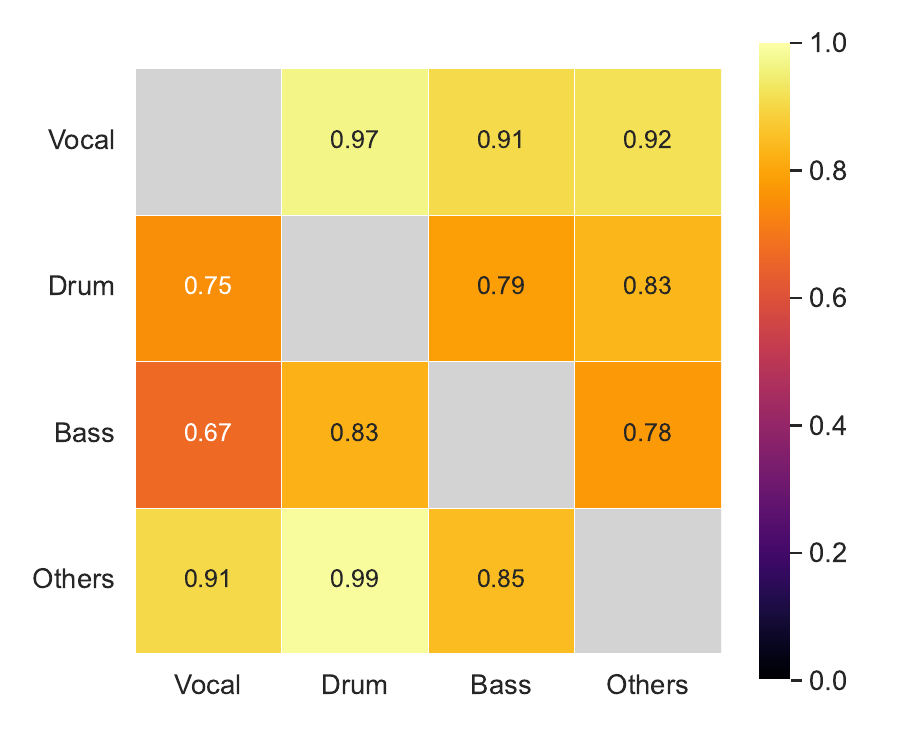}
    \caption{Sub \#2 (all-data)}
    \label{fig:subfig3}
\end{subfigure}

\vspace{0.1cm}

\begin{subfigure}{0.3\textwidth}
    \centering
    \includegraphics[width=\linewidth]{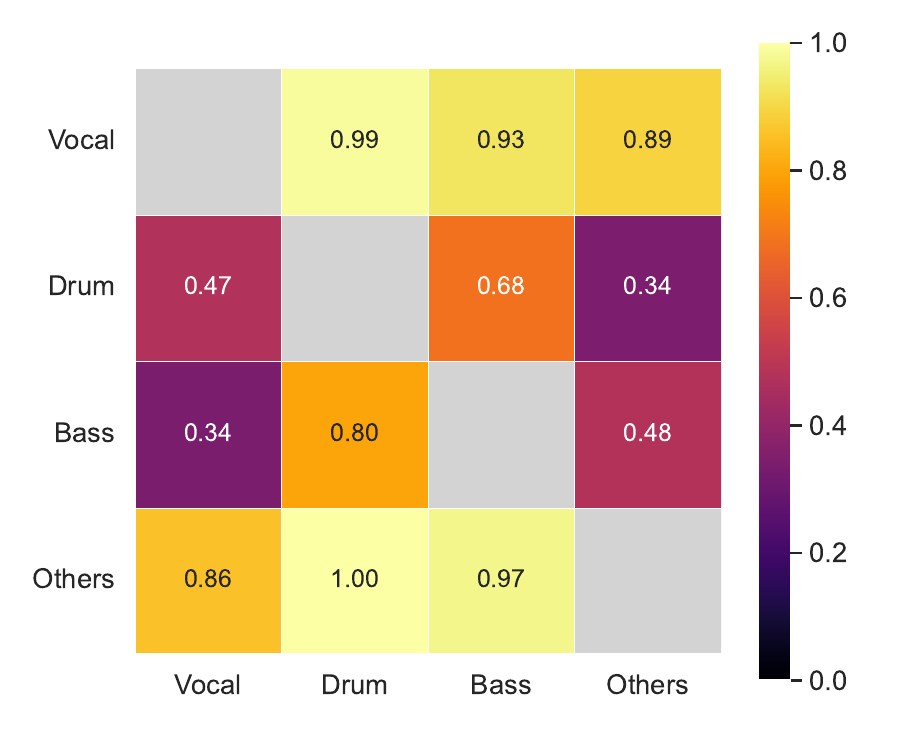}
    \caption{Sub \#3 (high-attention)}
    \label{fig:subfig4}
\end{subfigure}
\begin{subfigure}{0.3\textwidth}
    \centering
    \includegraphics[width=\linewidth]{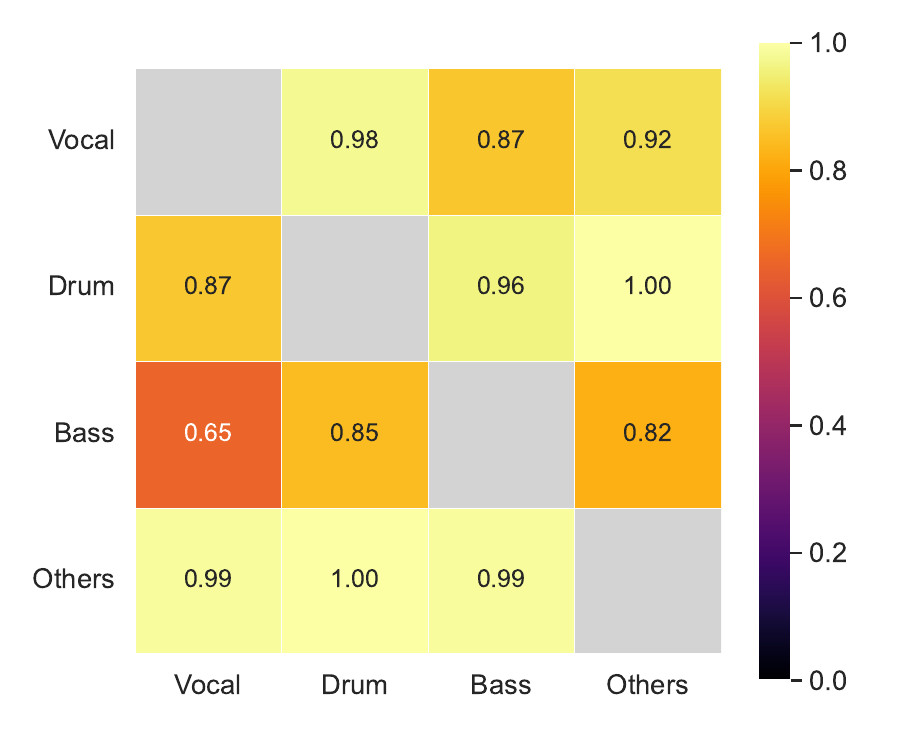}
    \caption{Sub \#7 (high-attention)}
    \label{fig:subfig5}
\end{subfigure}
\begin{subfigure}{0.3\textwidth}
    \centering
    \includegraphics[width=\linewidth]{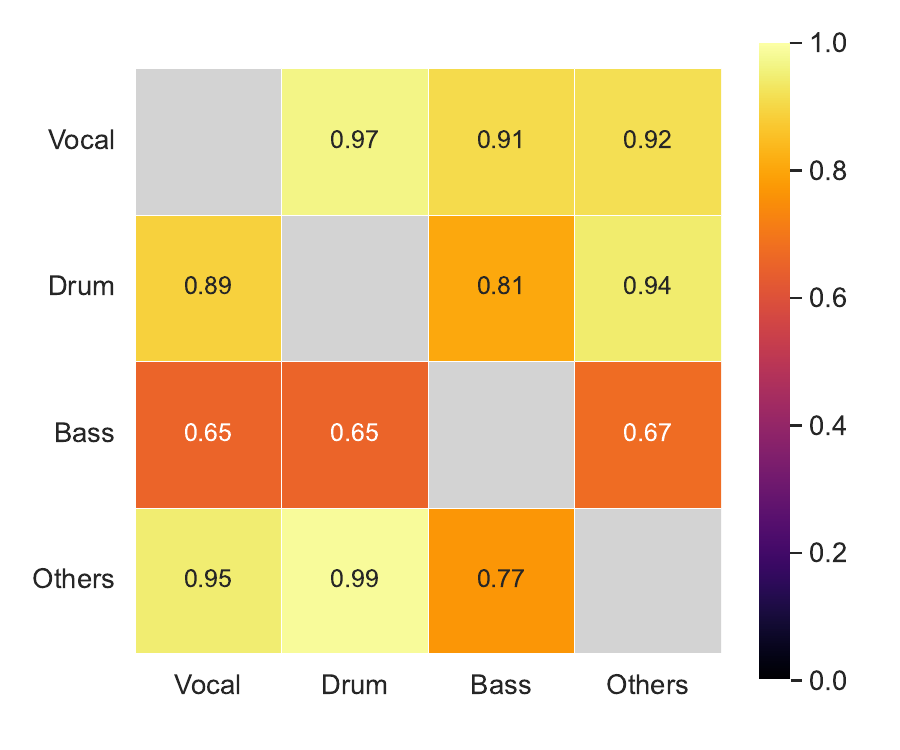}
    \caption{Sub \#2 (high-attention)}
    \label{fig:subfig6}
\end{subfigure}
\caption{Accuracy matrices across subjects.}
\label{fig:cross-subs_matrix}
\begin{minipage}{0.8\textwidth}
\normalsize
Panels (a)–(c) show pair-level accuracy for the three subjects under the all-data evaluation, while panels (d)–(f) correspond to the high-attention evaluation. Sub \#7 showed the best performance, followed by Sub \#2, whereas Sub \#3 exhibited the lowest performance.
\end{minipage}
\end{figure}

As shown in Figure~\ref{fig:cross-subs_matrix}, Sub \#7 yielded the best performance, with almost all task pairs around or exceeding 90\% accuracy, except for the \textit{Bass–Vocal} pair, which showed lower performance (75\% in the all-data evaluation and 65\% in the high-attention evaluation). Sub \#2 showed intermediate performance: the \textit{Drum} and \textit{Bass} pairs exhibited reduced accuracy but still remained around 80\%, except for the \textit{Bass–Vocal} pair, which reached 67\% in the all-data evaluation and 65\% in the high-attention evaluation, whereas the remaining pairs were around or above 90\%. In contrast, Sub \#3 showed the lowest performance, with \textit{Drum} and \textit{Bass} tasks performing particularly poorly—especially for the \textit{Drum–Vocal}, \textit{Bass–Vocal}, \textit{Drum–Others}, and \textit{Bass–Others} pairs—while the \textit{Vocal} and \textit{Others} tasks remained high, comparable to the other two subjects.

\subsection*{Previous paper comparison}
To compare our proposed model with the model from the most relevant study \cite{39}, we conducted a comparative analysis. In that work, only duos or trios from the MAD dataset were used, and thus their proposed model was designed to handle a maximum of three audio streams as input. However, since our experiment involves four streams (\textit{vocal}, \textit{drum}, \textit{bass}, and \textit{others}), we extended the model by adding a fourth stream with the same architecture as the original three. For EEG feature extraction methods such as CSP, we followed exactly the same procedure as described in the work\cite{39}. For comparison with our best model, \textbf{Model: all-0 ms}, the model from the previous study was also trained on all training data, with no time delay applied.

\subsubsection*{Within-subjects comparison}
\begin{figure}[H]
\centering
\includegraphics[width=0.6\linewidth]{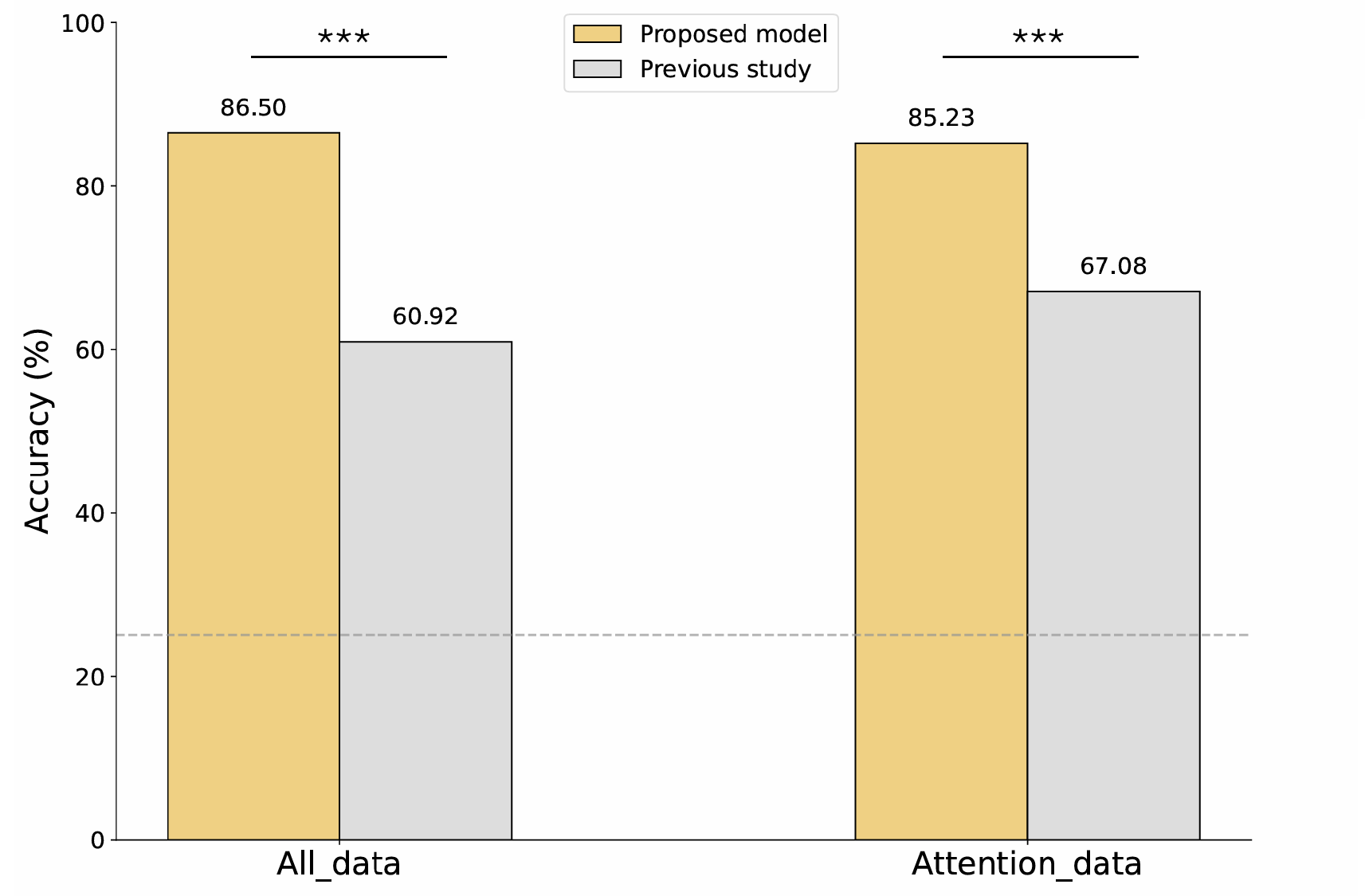}
\caption{Comparison of global accuracies.}
\label{fig:previous}
\begin{minipage}{0.8\textwidth}
\normalsize
The comparison was conducted for both all-data and high-attention data evaluations, with \textbf{Model: all-0 ms} shown in yellow and the previous study’s model in gray. The results suggest that our proposed model outperformed the previous study model, achieving significantly higher accuracy in both evaluations. $^{*} p < 0.05$, $^{**} p < 0.01$, $^{***} p < 0.001$ (McNemar test, compared with Previous study model).
\end{minipage}
\end{figure}

Figure \ref{fig:previous} shows the global accuracy comparison results, with the yellow bars representing our proposed model and the gray bars representing the model from the previous study\cite{39}. The x-axis is divided into two groups: all-data evaluation and high-attention data evaluation.

For both evaluation groups, our model showed significantly higher accuracy than the model from the previous study \cite{39}, with absolute differences of 25.58\% in the all-data evaluation ($p = 6.85 \times 10^{-27}$) and 18.15\% in the high-attention data evaluation ($p = 1.09 \times 10^{-13}$). Moreover, when comparing the accuracy difference between the two evaluation groups within each model, our model showed only a 1.27\% accuracy difference between two evaluations, whereas the previous study’s model exhibited a larger difference of 6.16\%. This indicates that our model may be less sensitive to attention-level differences.

\begin{figure}[H]
\centering
\includegraphics[width=0.9\linewidth]{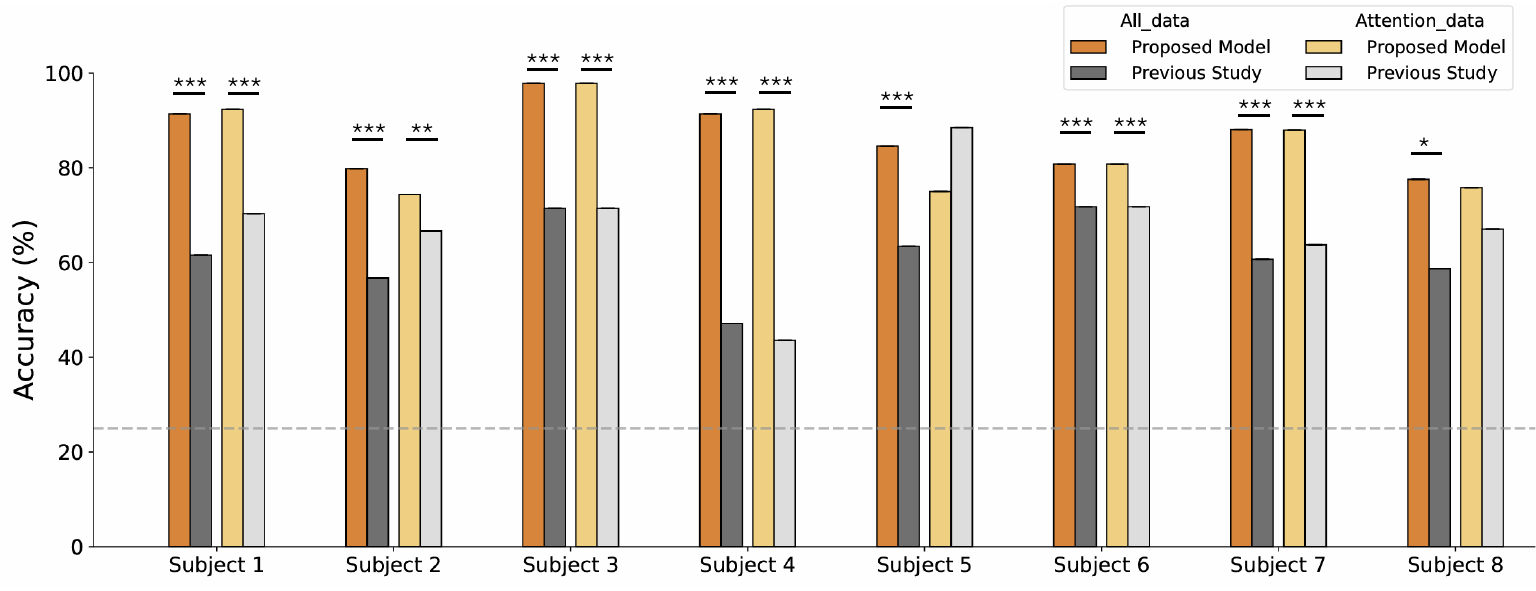}
\caption{Comparison of individual performances.}
\label{fig:previous_subs}
\begin{minipage}{0.8\textwidth}
\normalsize
The performance of all eight individuals is shown for both the all-data and high-attention evaluations, comparing with previous study's model. For most subjects, the observed effects reached statistical significance under both evaluation conditions. $^{*} p < 0.05$, $^{**} p < 0.01$, $^{***} p < 0.001$ (McNemar test, compared with Previous study model).
\end{minipage}
\end{figure}

We then compared individual performances between the two models. Figure \ref{fig:previous_subs} presents the results, where for each subject the two evaluation groups are shown separately: dark colors represent all-data evaluation (orange for the proposed model and black for the previous study model), while lighter colors represent high-attention data evaluation (yellow for the proposed model and gray for the previous study model). It should be noted that, due to slight differences in the number of trials and tasks for each individual, the average of the individual results does not exactly match the global results.

From these results, our proposed model significantly outperformed the model from the previous study for most subjects. The only exceptions were the high-attention evaluation for Sub \#5, where the previous study model achieved numerically higher accuracy but the difference was not significant ($p = 0.750$), and the high-attention evaluation for Sub \#8, where our model performed better numerically but again without a significant difference ($p = 0.291$). The most pronounced improvements for our model were observed in Sub \#4, with gains of 44.23\% in the all-data evaluation ($p = 1.96 \times 10^{-10}$) and 48.72\% in the high-attention evaluation ($p = 8.57 \times 10^{-9}$). Even for the lowest-performing subject, Sub \#2, our model yielded substantial gains, improving accuracy by 23.08\% in the all-data evaluation ($p = 7.43 \times 10^{-7}$) and by 7.69\% in the high-attention evaluation ($p = 1.73 \times 10^{-3}$).

Therefore, from both global accuracy and individual performance, our model achieved higher accuracy than the prior model and appeared less affected by low-attention segments in our tests. Even for those low-attended situation (participants feels difficult to find the target musical elements), our model can still effectively capture the features.

The specific values for each individual are shown in Table \ref{tab:previous_subs}.

\begin{table}[H]
\centering
\resizebox{\textwidth}{!}{
\begin{tabular}{lcccccccccccccccccccc}
\toprule
\diagbox{Model}{Subject} & sub1 & sub2 & sub3 & sub4 & sub5 & sub6 & sub7 & sub8 \\ 
\midrule
\textbf{Model: all-0 ms} (all-data)       & 0.9135$^{***}$ & 0.7981$^{***}$ & 0.9780$^{***}$ & 0.9135$^{***}$ & 0.8462$^{***}$ & 0.8077$^{***}$ & 0.8803$^{***}$ & 0.7755$^{*}$ \\
\textbf{Model: all-0 ms} (high-attention)  & 0.9231$^{***}$ & 0.7436$^{**}$ & 0.9780$^{***}$ & 0.9231$^{***}$ & 0.7500 & 0.8077$^{***}$ & 0.8791$^{***}$ & 0.7582 \\
Previous study model (all-data)            & 0.6154 & 0.5673 & 0.7143 & 0.4712 & 0.6346 & 0.7179 & 0.6068 & 0.5865 \\
Previous study model (high-attention)       & 0.7033 & 0.6667 & 0.7143 & 0.4359 & 0.8846 & 0.7179 & 0.6374 & 0.6703 \\
\bottomrule
\end{tabular}
}
\caption{Individual performance of \textbf{Model: all-0 ms}.}
\label{tab:previous_subs}
\centering
\begin{minipage}{0.8\textwidth}
\normalsize
This table presents the specific values for each individual across all models and evaluation groups, indicating higher accuracy and suggesting greater stability in attention state compared with the prior model\cite{39} in this setting. $^{*} p < 0.05$, $^{**} p < 0.01$, $^{***} p < 0.001$ (McNemar test, compared with Previous study model).
\end{minipage}
\end{table}

\subsubsection*{Cross-subjects comparison}
\begin{figure}[H]
\centering
\includegraphics[width=0.6\linewidth]{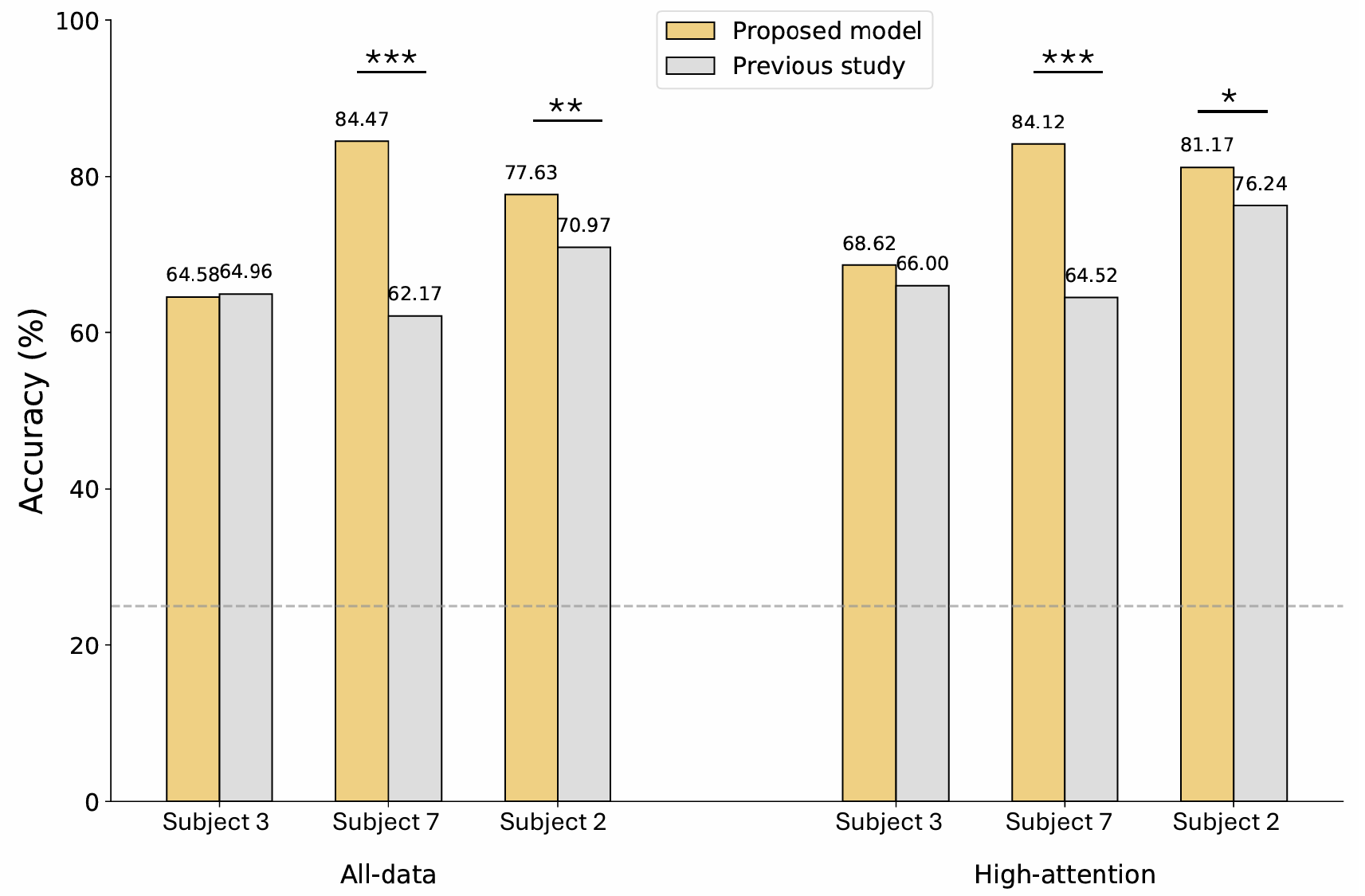}
\caption{Comparison of global accuracies in cross-subject evaluation.}
\label{fig:previous_cross}
\begin{minipage}{0.8\textwidth}
\normalsize
The comparison was performed for both all-data and high-attention evaluations, with \textbf{Model: all-0~ms} shown in yellow and the previous study’s model in gray. Except for sub \#3, our proposed model exceeded the previous study. $^{*} p < 0.05$, $^{**} p < 0.01$, $^{***} p < 0.001$ (McNemar test, compared with Previous study model).
\end{minipage}
\end{figure}

We further conducted a cross-subject comparison using the model reported in the previous study \cite{39}. Figure \ref{fig:previous_cross} presents the global accuracy results, with the x-axis divided into all-data evaluation and high-attention data evaluation. Yellow bars denote the performance of our proposed model, while gray bars correspond to the model from the previous study. 
Across both evaluation groups, our model achieved significantly higher accuracy than the previous study for sub \#7 and sub \#2. 
In particular, Sub \#7 exhibited substantial improvements of 22.3\% ($p = 1.33 \times 10^{-22}$) and 19.6\% ($p = 5.70 \times 10^{-15}$) in all-data and high-attention evaluations, respectively. Sub \#2 also showed pronounced gains of 6.66\% ($p = 1.43 \times 10^{-3}$) and 4.93\% ($p = 0.0288$) under the same conditions. However, no significant differences were observed for sub \#3 ($p = 0.956$ for all-data evaluation and $p = 0.282$ for high-attention data evaluation). 

Table \ref{tab:previous_cross} reports the detailed results for the two models. The proposed \textbf{Model: all-0~ms} achieved significantly higher performance compared with the previous study for sub \#7 and sub \#2.

\begin{table}[H]
\centering
\resizebox{0.7\textwidth}{!}{
\begin{tabular}{lcccccccccccccccccccc}
\toprule
\diagbox{Model}{Subject} & sub \#3 & sub \#7 & sub \#2 \\ 
\midrule
\textbf{Model: all-0 ms} (all-data)       & 0.6458 & 0.8447$^{***}$ & 0.7763$^{**}$ \\
\textbf{Model: all-0 ms} (high-attention)  & 0.6862 & 0.8412$^{***}$ & 0.8117$^{*}$ \\
Previous study model (all-data)            & 0.6496 & 0.6217 & 0.7097 \\
Previous study model (high-attention)       & 0.6600 & 0.6452 & 0.7624 \\
\bottomrule
\end{tabular}
}
\caption{Global accuracies in cross-subject evaluation.}
\label{tab:previous_cross}
\centering
\begin{minipage}{0.8\textwidth}
\normalsize
This table summarizes the results for three subjects across evaluation groups and shows that our model achieves significantly higher accuracies for Sub \#7 and Sub \#2. $^{*} p < 0.05$, $^{**} p < 0.01$, $^{***} p < 0.001$ (McNemar test, compared with Previous study model).
\end{minipage}
\end{table}

Taken together, the within- and cross-subject evaluations suggest that our model achieves higher accuracy than the prior model\cite{39} and exhibits greater stability across subjects and across different attention states. These findings point to the potential for real-world use, pending further validation.

\section*{Discussion}
\subsection*{Task-Specific Variability in Model Performance}
In the pair-level evaluations, the \textit{Bass} task yielded lower accuracy than the other tasks, particularly for \textbf{Model: attn-0 ms}. Two factors may account for this pattern:
(1) According to the participants, \textit{Bass} tasks were generally difficult to focus on, which was reflected in consistently low self-reported attention scores. Training the model only on high-attention trials may therefore be beneficial for tasks such as \textit{Vocal} and \textit{Drum}, where a sufficient number of high-attention trials are available. However, for \textit{Bass}, very few trials met the high-attention criterion, meaning that the model had limited opportunity to learn \textit{Bass}-specific features. As a result, the model performed poorly on the \textit{Bass} task despite being trained on nominally “high-quality” data.
(2) In some \textit{Bass}-related \textbf{incorrect} matches, large similarity differences were observed. Because the \textit{Bass} task is inherently difficult, the underlying signal quality is likely lower than for the other tasks, even when training is restricted to high-attention trials or when all data are used. In practice, this means that the training set for \textit{Bass} probably still contains many trials in which participants failed to attend appropriately, so the model is forced to learn from relatively noisy or unstable patterns. Consequently, the model tends to produce substantially deviating predictions for \textit{Bass}-related pairs.

For \textbf{Model: all-200 ms}, performance deteriorated most prominently for the \textit{Drum} task. This pattern may reflect the higher temporal sensitivity of drum-related processing. Unlike \textit{Vocal}, which unfolds as a relatively continuous auditory stream, \textit{Drum} events are expressed as discrete beats at specific time points. When an additional 200~ms delay is introduced between the EEG and the audio features, the analysis window is more likely to fall outside the critical time interval in which drum-related neural responses are present, leading to representations that lack drum information. By contrast, for more continuous stimuli such as \textit{Vocal}, even an imperfect temporal alignment may still capture vocal-related neural activity from adjacent time periods. Consequently, the \textit{Drum} task is more vulnerable to temporal misalignment, which may explain why the introduction of a 200~ms delay produced the largest performance decrement in this condition.

\subsection*{Reversal of Subject Ranking in the Cross-Subject Evaluation}
In the Cross-Subjects evaluation, the performance across the three subjects differed from that observed in the within-subject evaluation: Sub \#3, who had previously shown the highest accuracy, now exhibited the lowest performance, whereas Sub \#7 emerged as the best performer. One plausible explanation for this inversion is that the best-quality data from Sub \#3 were not included in the training set when Sub \#3 served as the test subject, thereby limiting the model’s ability to learn the most informative features from its best-quality data. Conversely, when evaluating Sub \#2, the training set did include the best data from Sub \#3, but the data quality of Sub \#2 itself may have been insufficient for reliable classification. As a result, the mid-level performer in the within-subject evaluation, Sub \#7, became the top performer in the cross-subject evaluation.

Another possible explanation relates to the specific neural characteristics of sub~\#3. In the within-subject evaluation, sub~\#3 exhibited extraordinarily high accuracy, whereas in the cross-subject evaluation very low accuracy was observed for sub~\#3 in the \textit{Drum} and \textit{Bass} tasks, indicating highly idiosyncratic brain features for this subject. Sub~\#3 may have contributed person-specific representations that the model could exploit when trained and tested on his or her own data, leading to inflated within-subject performance. However, when sub~\#3's data were excluded, no other subject exhibited comparable feature patterns, resulting in a degradation of cross-subject performance.

Such individual difference highlights the strong influence of both training-data composition and subject-specific data quality on cross-subject generalization. Future work should further investigate how training-data composition and subject-specific data quality interact to influence cross-subject generalization.

\subsection*{Time delays}
In this study, introducing a 200 ms delay between EEG and audio did not improve performance; instead, the best results were obtained with no additional delay. This contrasts with previous finding, which suggested that a 200 ms delay best reflects the latency between auditory input and neural responses\cite{predann}. One possible explanation is the use of a consumer-grade EEG device (Muse2) in our study, as opposed to research-grade equipment. Muse2 transmits EEG data to the MindMonitor application via Bluetooth, which likely introduces an intrinsic latency between the actual brain signals and the recorded data. Therefore, applying an additional artificial delay of 200 ms may have exceeded the optimal range, leading to degraded model performance. 
Nevertheless, the exact latency introduced by Muse2 and its Bluetooth transmission cannot be directly measured in our setup, and further investigation will be required to quantify this effect.

\subsection*{Brain regions}
In this study, we showed that decoding musical AAD was achievable using only four EEG channels (TP9, AF7, AF8, and TP10) located over frontal and temporal areas. Previous work has suggested that auditory attention decoding can achieve accuracy levels comparable to a 64-channel system even with only four electrodes (C4, C6, P9, and TP8)\cite{4channels}. However, that conclusion was drawn from a research-grade 64-channel setup, from which an optimal four-channel subset was identified through ablation analysis. By contrast, our study employed a consumer-grade device (Muse 2) with a fixed set of frontal–temporal electrodes (TP9, AF7, AF8, TP10). 

Our study is therefore distinct not only in using a small number of electrodes, but also in demonstrating effective decoding with electrode placements that are easy to apply and are already available in consumer-grade devices. Prior studies have reported hemispheric dominance in speech and auditory perception, with substantial inter-individual variability\cite{dominance_1,dominance_2,auditory_dominance}. Other work has highlighted distinct hemispheric roles, such as a greater contribution of the left hemisphere to temporal processing and of the right hemisphere to timbre processing\cite{right_hemiephere}. Our results indicate that a small set of frontal–temporal electrodes can provide effective decoding of musical attention. Future work may systematically explore how specific electrode placements and hemisphere-focused configurations affect decoding efficiency and the robustness of musical AAD.

\subsection*{Challenges in Generalization to New Subjects and Songs}
In the present study, the evaluation focused on two generalization settings: (1) predicting attentional focus on new songs within the same participant, and (2) predicting attentional focus for a new participant on the same songs (leave-one-subject-out). While these settings suggest robustness within controlled conditions, they do not fully capture scenarios where both the listener and the songs are novel.

Future research could therefore explore more challenging generalization conditions, such as evaluating attentional focus when an existing user listens to entirely new songs, or when a new user listens to songs not present in the training set. Moreover, given the increasing accessibility of consumer-grade EEG devices, it would be feasible to collect large-scale data in more naturalistic environments. 

\section*{Methods}
\subsection*{Dataset}
The experimental session lasted approximately 45 minutes and consisted of 63 trials; each derived from a different song selected from the MUSDB18 dataset\cite{musdb18}. 
A 15-second excerpt containing the target element was randomly selected as the stimulus for one trial, and the same excerpts were presented to all participants.
The presence of the target element in each excerpt was confirmed using librosa.trim \cite{librosa} to ensure task validity.

To compare attention to vocals versus other instruments, the trials were distributed as follows: 31 trials targeting vocals, 11 targeting drums, 11 targeting bass, and 10 targeting other instruments. Musical genres were randomly sampled to ensure diversity and to approximate real-world listening conditions. The MUSDB18 dataset includes a broad range of musical genres—such as pop, rock, hip-hop, jazz, and electronic—ensuring that the selected stimuli reflect the diversity of real-world music and support generalization across musical styles. To avoid any systematic association between genre and attentional target, both the target element and the musical genre were independently and fully randomized across trials. 

For the within-subject evaluation on new songs, each participant’s data were randomly divided into a ratio of 8:1:1 for training, validation, and testing, ensuring that all test data consisted of songs not previously encountered by the model. For the cross-subject evaluation on new participants, a leave-one-subject-out approach was adopted: one subject was held out as the test set, while the data from the remaining subjects were divided into training and validation sets with an 8:2 ratio within each subject.

\subsection*{Experiment setup and protocol}
\begin{figure}[H]
\centering
\begin{subfigure}{0.3\textwidth}
    \centering
    \includegraphics[width=\linewidth]{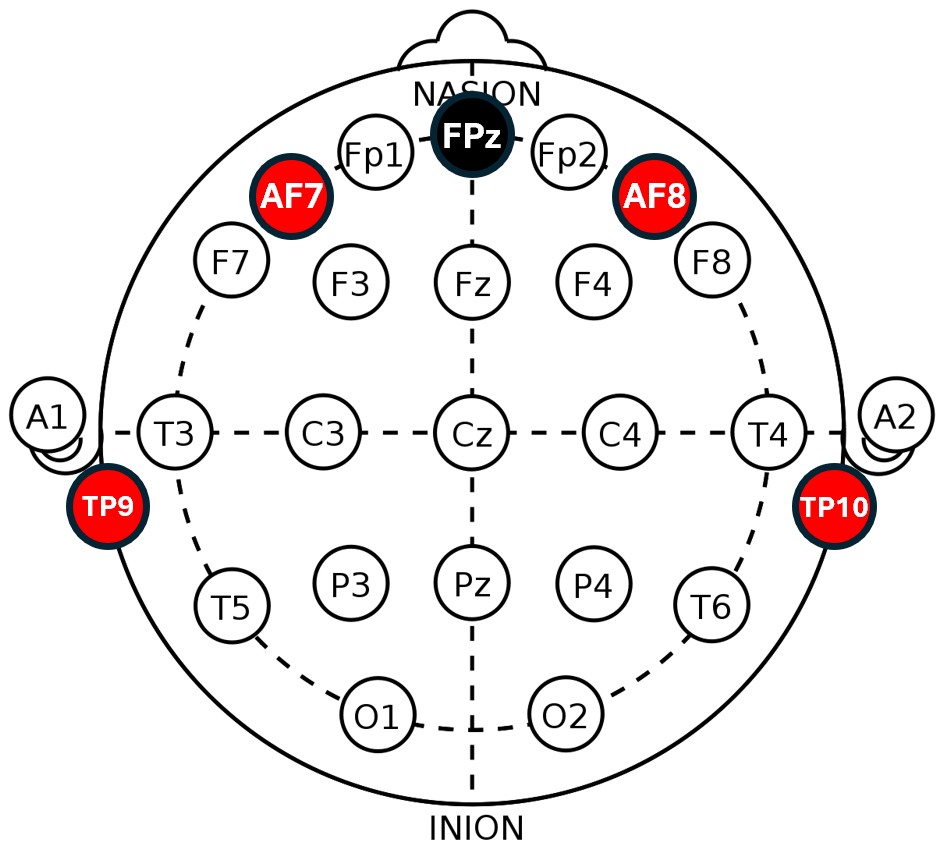}
    \caption{Electrode placement.}
    \label{fig:electrodes}
\end{subfigure}
\hfill
\begin{subfigure}{0.6\textwidth}
    \centering
    \includegraphics[width=\linewidth]{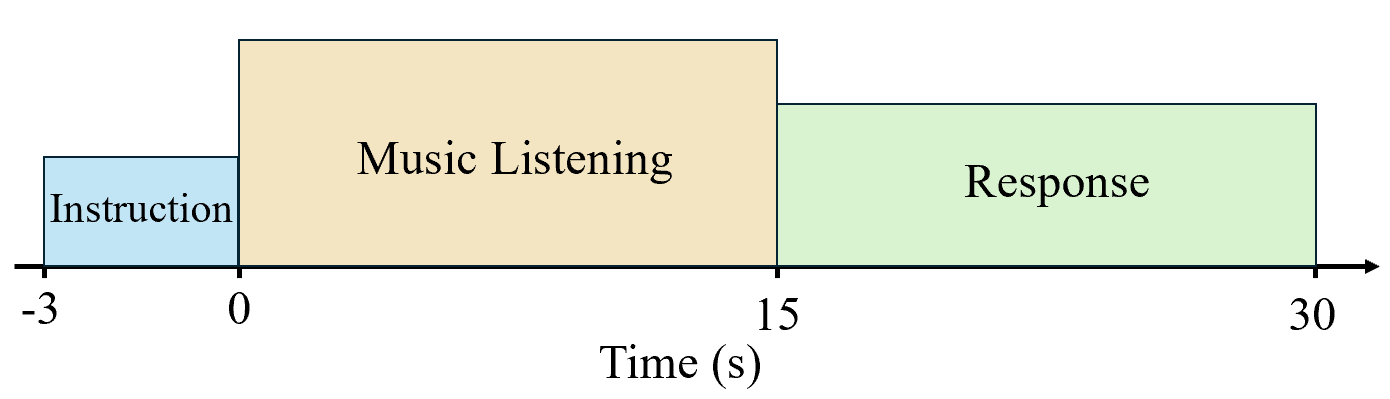}
    \caption{Experimental protocol.}
    \label{fig:protocol}
\end{subfigure}
\caption{Overview of the experimental setup.}
\label{fig:combined_protocol}
\begin{minipage}{0.5\textwidth}
\normalsize
\centering
(a) Electrode placement. (b) Experimental protocol.
\end{minipage}
\end{figure}

Brain activity was recorded using the Muse 2 headband\cite{muse2}, a portable EEG device designed for practical and real-world applications. EEG signals were sampled at 256 Hz using four dry electrodes located at TP9, AF7, AF8, and TP10, with the reference electrode at FPz as shown in Figure \ref{fig:electrodes}.

The experiment was carried out in a quiet room with minimal external disturbances to emulate a realistic setting for EEG measurement. Participants were seated comfortably and instructed to minimize body movement throughout the session to reduce motion artifacts. All audio stimuli were presented at a sampling rate of 44,100 Hz through Sony inner earphone MDR-EX15LP. This stereo presentation replicated natural music listening conditions while avoiding any artificial spatial cues that could guide participants' attention.

Figure \ref{fig:protocol} illustrate the protocol of the experiment. Prior to the start of the experiment, participants were informed of the overall procedure and completed a brief practice session consisting of three example trials, allowing them to become familiar with the task structure and types of auditory cues.
Each experimental trial consisted of three phases: a 3-second instruction phase, a 15-second auditory task phase, and a 15-second response phase. During the Instruction phase, participants were visually cued (e.g., "Focus on vocals") via text presented at the center of the screen. In the Music Listening phase, they listened to a 15-second segment of music and were instructed to focus their attention on the specified elements (vocals, drums, bass, or others). To minimize eye movement and visual distraction, a fixation cross was displayed at the center of the screen throughout the task phase. In the Response phase, participants rated their attentional success on a 5-point Likert scale (1 = not at all, 5 = fully attentive). These subjective reports were later used as a supplementary behavioral measure. 
The protocol was implemented in surveyjs\cite{surveyjs}, and a schematic of the trial structure is shown in Figure\ref{fig:protocol}.

\subsection*{Participants}
Nine participants (age range: 25-53 years; 7 males, 2 females) were initially recruited for the experiment. Data from one male participant was excluded due to poor signal quality, resulting in a final sample of eight individuals (M = 37.25 years, SD = 10.49). All participants self-reported normal hearing and no history of neurological or musculoskeletal disorders. All participants gave their informed consent prior to the experiment, and the study was conducted in accordance with general ethical guidelines.

\subsection*{EEG Preprocessing}

Following Défossez et al. \cite{Defossez2022}, EEG signals were normalized using the \textit{RobustScaler}, followed by a value clamping operation to improve numerical stability. The \textit{RobustScaler} computes the median and interquartile range (IQR) for each channel and scales data accordingly, effectively reducing the influence of outliers—a common issue in EEG signals due to sporadic noise and artifacts. After scaling, values were clamped within a fixed range of ±20: values below -20 were set to -20, and those above +20 were set to +20. This two-step normalization was applied channel-wise. Given the sampling rates of EEG (256 Hz) and audio (44,100 Hz), the signals were zero-padded to fixed lengths of $3^7$ and $3^{11}$, respectively, corresponding to 3-s segments, to facilitate compatibility with the model architecture.

\subsection*{Model Architecture and Loss}
A 2D convolutional neural network (2D-CNN) was used to extract features from EEG data. We adopted a contrastive learning approach to align EEG representations with embeddings of individual musical components (vocals, drums, bass, and others), following our previous work \cite{predann}. For each EEG segment, similarity scores were computed with respect to the candidate musical embeddings, and the element with the highest similarity was identified as the attended source. This architecture differs from our previous work \cite{predann}, which incorporated both classification and contrastive objectives; here, we focus solely on the contrastive alignment to directly decode attentional targets at the element level.
This design follows our earlier finding that learning shared representations between cortical activity and ANN-based audio embeddings enables robust decoding, even in noisy and artifact-prone EEG conditions. Therefore, no additional artifact rejection or filtering was applied to the raw EEG. 

\begin{figure}[H]
\centering
\includegraphics[width=0.8\textwidth]{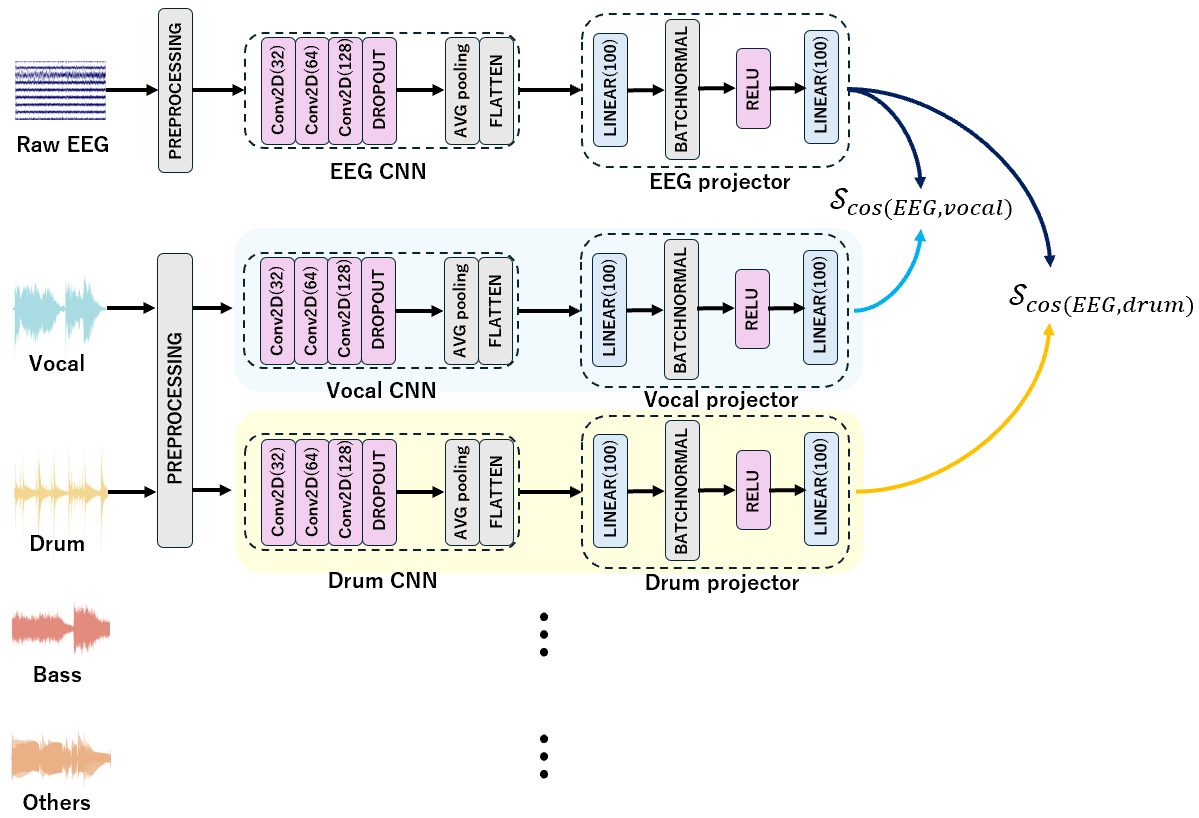}
\caption{The proposed model.}
\label{fig:proposed model}
\centering
\begin{minipage}{0.8\textwidth}
\normalsize
Each stream (one EEG and four audio inputs) was processed independently by a 2D CNN. For each audio stream, the cosine similarity with the EEG signal was computed. The audio stream with the highest similarity score was selected as the final output, indicating which element (vocal, drum, bass, or others) the EEG signal attends to.
\end{minipage}
\end{figure}

For the loss calculation, we denote the EEG feature and four audio-stem features of the $i$-th sample as
\[
\{(e_i, v_i, d_i, b_i, o_i)\}_{i=1}^{B}, \quad
e_i, v_i, d_i, b_i, o_i \in \mathbb{R}^D.
\]
For each of the four tasks, the corresponding audio feature is defined as
$a_i^{(v)} = v_i$, $a_i^{(d)} = d_i$, $a_i^{(b)} = b_i$, and $a_i^{(o)} = o_i$.
The dataset is organized into four task-specific groups, and the loss for each task is computed independently. Let $\{\mathbf{z}_i^{E}\}_{i=1}^{B}$ and $\{\mathbf{z}_i^{A_m}\}_{i=1}^{B}$ be the outputs of the EEG projector and the audio projector of modality $m\in\{v,d,b,o\}$, respectively, in a mini-batch of size $B$. For task $t\in\{v,d,b,o\}$, the $i$-th EEG is paired positively with the audio of the same index from modality $t$,forming the positive pair $(\mathbf{z}_i^{E},\,\mathbf{z}_i^{A_t})$. The task-specific loss is defined as:
\begin{equation}
\mathcal{L}_t
= -\frac{1}{B}\sum_{i=1}^{B}
\log
\frac{\exp\!\big(\operatorname{sim}(\mathbf{z}_i^{E},\,\mathbf{z}_i^{A_t})\big)}
{\displaystyle\sum_{m\in\{v,d,b,o\}}\sum_{j=1}^{B}
\exp\!\big(\operatorname{sim}(\mathbf{z}_i^{E},\,\mathbf{z}_j^{A_m})\big)}.
\label{eq:task-loss}
\end{equation}
where $\texttt{sim}(\cdot,\cdot)$ denotes cosine similarity and $\tau$ is the temperature parameter. The denominator includes all \textit{4B} candidates, ensuring that each EEG representation is contrasted against the full set of audio embeddings across all modalities. 

The loss framework optimizes mutual information between EEG and music representations rather than likelihood-based conditional modeling.
In practice, we adopt the InfoNCE loss, which serves as a lower-bound estimator of negative mutual information and was originally introduced in Contrastive Predictive Coding for speech and image \cite{Oord2018RepresentationLW}.
Cross-modal extensions of this objective were first demonstrated in ConVIRT\cite{pmlr-v182-zhang22a} and subsequently popularized by CLIP\cite{pmlr-v139-radford21a}. 

The overall training objective is obtained by taking an unbiased mean across the four task-specific losses:
\begin{equation}
\mathcal{L}_{\text{final}}
= \frac{1}{4}\big(
\mathcal{L}_v
+ \mathcal{L}_d
+ \mathcal{L}_b
+ \mathcal{L}_o
\big).
\end{equation}

\subsection*{Data Augmentation}

To augment training data, a sliding window approach was applied to the EEG recordings. For training and validation, a window size of 1280 samples (5 seconds) and a stride of 256 samples (1 second) were used. Within each 5-second window, a 3-second segment was randomly sampled as input to the model. To introduce further augmentation, the initial EEG offset was randomly sampled from 0 to 255 in each epoch, with the corresponding audio segment shifted accordingly to maintain alignment.

For the testing set, no randomization was applied in order to simulate continuous real-world listening conditions. Each test sequence was segmented using a fixed window size of 768 samples (3 seconds) with the same stride of 256 samples.

\subsection*{Training and evaluation}
Training and validation were conducted using the InfoNCE contrastive learning function following the PredANN framework \cite{predann}. 
For pairs definition, if the first trail is the \textit{vocal} task, the EEG data of the first trial is denoted as EEG\_vocal\_1, and the corresponding audio stimuli include audio\_vocal\_1, audio\_drum\_1, audio\_bass\_1, and audio\_others\_1.
Similarly, if the second trial corresponds to the \textit{drum} task, the EEG data is denoted as EEG\_drum\_2, and its associated audio stimuli are audio\_vocal\_2, audio\_drum\_2, audio\_bass\_2, and audio\_others\_2.
Positive pairs were defined as temporally aligned EEG–audio pairs from the same trial, considered in both EEG$\rightarrow$audio and audio$\rightarrow$EEG directions. For example, EEG\_vocal\_1 was paired with audio\_vocal\_1, and audio\_drum\_2 with EEG\_drum\_2. 
Negative pairs were constructed from three categories, again in both EEG$\rightarrow$audio and audio$\rightarrow$EEG directions:
\begin{enumerate}[label=(\roman*)]
    \item \textbf{Within-trial cross-task negatives:} EEG signals paired with non-attended audio components from the same trial (e.g., EEG\_vocal\_1 with audio\_drum\_1).
    \item \textbf{Across-trial same-task negatives:} EEG signals paired with audio segments of the same task type but from different trials (e.g., EEG\_vocal\_1 with audio\_vocal\_2).
    \item \textbf{Across-trial cross-task negatives:} EEG signals paired with audio segments of different task types from other trials (e.g., EEG\_vocal\_1 with audio\_drum\_2).
\end{enumerate}
The InfoNCE objective encourages the model to maximize the similarity of each positive EEG–audio pair while simultaneously minimizing similarity with all negative pairs.
For a batch of size $4N$, each trial produces one positive pair similarity and $4N-1$ negative pairs, which are used together to compute the loss.
This loss was used to supervise the model under three training strategies: (1) using all trials, (2) using only high-attention trials (with self-reported attention scores equal to or above 4), and (3) applying a 200 ms temporal delay between the audio and EEG input streams. The training process was conducted for a maximum of 1000 epochs, employing early stopping with a patience of 10 epochs to mitigate overfitting, while enforcing a minimum of 50 epochs to prevent premature termination.

In the evaluation, we assess performance in terms of accuracy, defined as the proportion of instances in which the instrument attended to by the subject is correctly identified. Our accuracy evaluation proceeded in three steps:

\textbf{1. 4×4 matrix of pair-level accuracy:}  
   We constructed a $4 \times 4$ matrix in which each off-diagonal element $(i, j)$ represented the proportion of cases where the similarity between EEG$_i$ and audio$_i$ (EEG$_i$ was recorded when the subject attended to the music component audio$_i$) exceeded the similarity between EEG$_i$ and audio$_j$ (audio$_j$ is included as a stem in the same music piece as audio$_i$). Diagonal entries were left undefined. Each value in this matrix reflects how well the model distinguishes the correct attended musical element for a given EEG pattern relative to distractors from other audio stems.

\textbf{2. 4×1 vector of task-level accuracy:} 
   For each EEG source (vocal, drum, bass, others), we computed the proportion of comparisons in which the similarity between EEG$_i$ and audio$_i$ (EEG$_i$ was recorded when the subject attended to the music component audio$_i$) exceeded all the similarities between EEG$_i$ and audio$_j$ (audio$_j$ is included as a stem in the same music piece as audio$_i$). This resulted in a 4×1 vector, where each value reflects the model’s overall ability within each task.

\textbf{3. Global accuracy:} 
   The overall performance of the model was quantified by a single scalar accuracy reflecting its ability to match EEG signals to their correct attention context.

Next, we categorized all EEG–audio comparisons into two groups based on performance:

- \textbf{Group 1 (Correctly matched):} Cases where the EEG–audio positive pair had higher similarity than all corresponding negative pairs.

- \textbf{Group 2 (Incorrectly matched):} Cases where the positive pair had lower similarity than at least one negative comparison.

4×4 matrix of pair-level similarity difference
4×1 vector of task-level similarity difference
Global similarity difference

For each group, we calculated the similarity difference between the positive and each negative pair. This analysis was again summarized at three levels:

\textbf{1. 4×4 matrix of pair-level similarity difference}, where each entry indicates how much more (or less) similar the EEG–audio positive pair was compared to a specific negative pairing;

\textbf{2. 4×1 vector of task-level similarity difference} of average similarity differences for each task;

\textbf{3. Global similarity difference}, capturing the overall contrast between positive and negative pairs across all tasks.

Together, these analyses provided a comprehensive evaluation framework for measuring how effectively the model captured task-specific EEG–audio correspondence and attentional consistency.

\section*{Conclusion}
In this study, we present the first attempt in the field of auditory attention decoding (AAD) to use naturalistic, studio-produced songs together with a lightweight, consumer-grade EEG device equipped with only four electrodes. Through this novel approach, we addressed two primary objectives. From a neuroscientific perspective, we investigated whether decoding can be achieved using real music that users can naturally enjoy without burden, thereby providing new insights into the neural mechanisms of musical attention. From an application perspective, we demonstrated that collecting data from diverse participants in naturalistic conditions can not only advance scientific understanding but also lay the foundation for practical real-world applications.

Our results show that the proposed model outperforms previous methods not only within individual subjects but also generalizes well to unseen subjects, highlighting the feasibility of future real-world application development. In addition, our dataset encompasses a wide range of musical genres, bridging a critical gap with prior studies that have predominantly relied on artificial stimuli such as simple tones or spatialized sounds. This diversity underscores the robustness of our approach and its applicability to realistic listening environments.

Finally, the use of the Muse2 headset confirmed that reliable decoding of attentional focus is achievable with only four EEG channels, offering neuroscientific evidence that temporal and frontal brain regions may play a dominant role in musical attention. Taken together, these findings establish an important milestone in the field, marking the first demonstration that music AAD can be realized under naturalistic conditions with minimal hardware, thereby paving the way for both future research and practical applications in real-world neural decoding of music.

\bibliography{references}

\section*{Figure Legends}
\textbf{Figure 1. Global accuracy.}  
The figure illustrates the global accuracy among the three baseline models.

\textbf{Figure 2. Global similarity differences.}  
The figure illustrates the global similarity differences among the three baseline models.

\textbf{Figure 3. Task-level accuracy vectors.}  
The figure shows the task-level accuracy vectors for the baseline models.

\textbf{Figure 4. Task-level similarity difference vectors.}  
The figure shows the task-level similarity difference vectors for the baseline models.

\textbf{Figure 5. Accuracy matrices.}  
The figure shows the pair-level accuracy results for the baseline models.

\textbf{Figure 6. Similarity difference matrices (all-data).}  
The pair-level similarity difference results are divided into \textbf{Correctly matched} and \textbf{Incorrectly matched} groups for the baseline models under the all-data evaluation setting.

\textbf{Figure 7. Similarity difference matrices (high-attention).}  
The pair-level similarity difference results are divided into \textbf{Correctly matched} and \textbf{Incorrectly matched} groups for the baseline models under the high-attention data evaluation setting.

\textbf{Figure 8. Song-level performance within each subject.}  
Three typical subjects are selected: the highest (Sub \#3), middle (Sub \#7), and lowest (Sub \#2). Results of the specific task evaluation are shown for each subject.

\textbf{Figure 9. Task-level evaluation across subjects.}  
The figure shows the task-level cross-subject evaluation results for the same three subjects shown in Figure \ref{fig:specific}.
 
\textbf{Figure 10. Task-level evaluation across subjects.}  
The figure shows the pair-level cross-subject evaluation results for the same three subjects as in Figure\ref{fig:specific}.

\textbf{Figure 11. Comparison of global accuracies.}  
The figure shows the results compared with the model proposed in a previous study \cite{39} in terms of global accuracy.

\textbf{Figure 12. Comparison of individual performances.}  
The figure shows the results compared with the previous study model\cite{39} for each individual.

\textbf{Figure 13. Comparison of global accuracies in cross-subject evaluation.}  
The figure shows the cross-subject results compared with the previous study model\cite{39} for same 3 subjects in \ref{fig:specific}.

\textbf{Figure 14. Overview of the experimental setup.}  
The figure shows (a) Electrode placement. (b) Experimental protocol.

\textbf{Figure 15. The proposed model.}  
The figure illustrates the architecture of the proposed model.

\section*{Author contributions statement}
T.A. and Z.Z. contributed equally to this work.
T.A., Z.Z., and T.N. conceived the study and designed the overall framework, methodology, and experiments.
Z.Z. and T.N. implemented the code and, together with T.Y., conducted the experiments.
T.A., Z.Z. and N.P. analyzed and discussed the results.
Z.Z. and T.A. drafted the main manuscript and prepared the tables and figures.
Z.Z., T.A. and S.M. organized the code.
N.P., T.A., and Z.Z. critically reviewed and revised the manuscript.
N.P. and T.A. advised the research, and N.P. organized the research project.

\end{document}